 \documentclass[a4paper,fleqn]{cas-dc}

\usepackage[numbers]{natbib}
\usepackage{float}
\usepackage{amsmath}
\usepackage{physics}
\usepackage{amssymb}
\usepackage{times}
\usepackage{mathtools}
\usepackage{rotating}
\usepackage{siunitx}
\usepackage{tabularx}
\usepackage{placeins}
\usepackage{booktabs}
\usepackage{makecell}
\usepackage{longtable}

\def\tsc#1{\csdef{#1}{\textsc{\lowercase{#1}}\xspace}}
\tsc{WGM}
\tsc{QE}
\tsc{EP}
\tsc{PMS}
\tsc{BEC}
\tsc{DE}
\makeatletter
\def\fps@table{!hbp}
\makeatother

\begin{document}
\let\WriteBookmarks\relax
\def\floatpagepagefraction{1}
\def\textpagefraction{.001}
\interfootnotelinepenalty=10000

% Short title
\shorttitle{A quantitative phase-field model for grain boundary trapping and diffusion of hydrogen}

% Short author
\shortauthors{Feyen et~al.}

% Main title of the paper
\title[mode = title]{A quantitative phase-field model for grain boundary trapping and diffusion of hydrogen}

% Title footnote 1.
% eg: \tnotetext[1]{Title footnote text}
% \tnotetext[<tnote number>]{<tnote text>} 
%\tnotetext[1]{This document is the results of the research
%  project funded by the Research Foundation - Flanders (FWO).}

%\tnotetext[2]{The second title footnote which is a longer text matter
%  to fill through the whole text width and overflow into
%  another line in the footnotes area of the first page.}

% First author
%
% Options: Use if required
% eg: \author[1,3]{Author Name}[type=editor,
%    style=chinese,
%    auid=000,
%    bioid=1,
%    prefix=Sir,
%    orcid=0000-0000-0000-0000,
%    facebook=<facebook id>,
%    twitter=<twitter id>,
%    linkedin=<linkedin id>,
%    gplus=<gplus id>]
\author[1]{Vincent Feyen}[type=editor,
            auid=000,bioid=1,
            orcid=0000-0002-6079-4477]
\cormark[1] %Corresponding author indication
\fnmark[1] % Footnote of the first author
\ead{vincent.feyen@kuleuven.be} % Email id of the first author
\credit{Conceptualization, Software, Data curation, Methodology, Software, Writing - Original Draft} 

\author[1,2]{Martin Diehl}
\credit{Writing - Review \& Editing, Supervision}

\author[1]{Nele Moelans}
\credit{Writing - Review \& Editing, Supervision}

% Address/affiliation
\affiliation[1]{organization={Department of Materials Engineering, Faculty of Engineering, KU Leuven},
  addressline={Kasteelpark Arenberg 44, bus 2450}, 
  city={Leuven},
  % citysep={}, % Uncomment if no comma needed between city and postcode
  postcode={3001}, 
  % state={},
  country={Belgium}}

\affiliation[2]{organization={Department of Computer Science, Faculty of Engineering Science, KU Leuven},
  addressline={Celestijnenlaan 200A, bus 2402}, 
  city={Leuven},
  % citysep={}, % Uncomment if no comma needed between city and postcode
  postcode={3001}, 
  % state={},
  country={Belgium}}

% Here goes the abstract - max 200 words

% \begin{abstract}
% test
% \end{abstract}

\begin{abstract} 
Grain boundaries influence hydrogen transport in polycrystalline metals by acting as both trapping sites and interconnected diffusion pathways. Conventional interpretations of thermal desorption spectroscopy (TDS) and permeation experiments often treat traps as isolated defects, neglecting grain boundary connectivity and potentially misinterpreting experimental data. Here, we develop a quantitative phase-field model for hydrogen diffusion, grain boundary trapping, and grain boundary-assisted transport. The formulation preserves the physical grain boundary volume independently of the numerical interface thickness and accounts for anisotropic diffusion parallel and perpendicular to grain boundaries. Benchmark simulations verify quantitative behavior under interface upscaling. The model shows that grain boundary diffusion can significantly shift TDS peaks, making Kissinger-type analyses unreliable when grain boundary transport is significant. Effective diffusion coefficients are also shown to depend strongly on grain size, trapping free energy, temperature, and grain boundary mobility. These results highlight the coupled role of trapping thermodynamics, bulk diffusion, and grain-boundary transport.
\end{abstract}

% Use if graphical abstract is present
% \begin{graphicalabstract}
% \includegraphics{figs/grabs.pdf}
% \end{graphicalabstract}

% % Research highlights
% \begin{highlights}
% \item Research highlights item 1
% \item Research highlights item 2
% \item Research highlights item 3
% \end{highlights}

% Keywords
% Each keyword is seperated by \sep

\maketitle

%\tableofcontents

\section{Introduction}
\label{sec:Introduction}
Hydrogen produced from renewable electricity is widely regarded as a key energy carrier for decarbonizing industrial processes, energy storage, and transportation. Its deployment, however, places stringent demands on the safety and durability of structural materials used in hydrogen production, storage, and transport systems. Because hydrogen has a low ignition energy, a wide flammability range, and is commonly stored and transported at high pressures to achieve practically useful energy densities, the structural integrity of pipelines, storage vessels, and pressure-bearing components exposed to hydrogen environments is of central importance.

A major materials-related challenge in this context is hydrogen embrittlement. Owing to its small atomic size, hydrogen readily penetrates most metals and diffuses rapidly through their crystal lattices, even at ambient temperatures. Once absorbed, hydrogen can severely degrade mechanical properties, leading to reduced ductility and premature brittle failure. Despite decades of research, hydrogen embrittlement remains only partially understood and is known to depend sensitively on alloy composition, temperature, mechanical loading conditions, and hydrogen exposure history \cite{Chen2025, Cotterill1961, Lee2016}. Comprehensive overviews of proposed embrittlement mechanisms are available in the literature \cite{Koyama2017, Hirth1980, Yu2024, Chen2025, Negi2024, Li2020}. A common feature among these mechanisms is the central role of hydrogen diffusion and redistribution within the material \cite{Lee2016, Yu2024, Cotterill1961}.

For many structural alloys of interest to hydrogen energy applications, such as steels, the equilibrium solubility of hydrogen in the bulk lattice is very low \cite{Cotterill1961}. Nevertheless, hydrogen embrittlement can occur even at low average concentrations, i.e., a few wt. ppm of H~\cite{Cotterill1961}, because hydrogen preferentially accumulates at microstructural defects, including grain boundaries (GBs), dislocations, vacancies, carbides, and other second-phase features \cite{Koyama2017, Yu2024}. These defects act as trapping sites where local hydrogen concentrations can be orders of magnitude higher than in the surrounding lattice, with atomistic studies reporting enrichment factors exceeding \(300\) at some grain boundaries~\cite{Smirnova2023}, thereby strongly influencing both the total hydrogen content and the effective transport behavior \cite{Yu2024, Oriani1970}. Among these defects, GBs are of particular importance because they are both ubiquitous in engineering alloys and interconnected at the macroscopic scale, allowing them to act not only as trapping sites but also as long-range transport pathways. Both aspects mainly depend on the atomic structure and composition of the GBs \cite{Brass1996, Iwaoka2016, Oudriss2012, McEniry2017}. This dual role was demonstrated for pure nickel by Oudriss et al.~\cite{Oudriss2012}, who showed that high-angle GBs can act as fast diffusion pathways, whereas low-angle GBs are dominated by trapping. This conclusion was also supported by ab-initio modeling \cite{Zhou2019}.

Two experimental techniques are widely used to characterize hydrogen transport and trapping in metals: thermal desorption spectroscopy (TDS) and hydrogen permeation measurements. In a typical TDS experiment, hydrogen-charged specimens are heated at a controlled rate while the evaporated hydrogen is monitored, producing a desorption spectrum that reflects a convolution of trapping thermodynamics and transport kinetics \cite{Verbeken2012, vonZeppelin2003}. In permeation experiments, the time evolution of hydrogen flux through a sample is measured and subsequently interpreted in terms of an effective diffusion coefficient \cite{ASTM_G0148_97R18}. In both cases, interpretation remains challenging because multiple trapping sites and diffusion pathways may contribute simultaneously to the measured signal. Classical analysis methods, such as Gaussian peak fitting and Kissinger-type analysis, used for TDS analysis or effective diffusion expressions derived for isolated traps based on permeation experiments, do not explicitly account for the spatial distribution and connectivity of microstructural defects \cite{Blaine2012, Choo1982, Silva2024, Drexler2021, Oriani1970}. As a result, the extracted trapping energies, trap densities, or diffusion coefficients may conflate thermodynamic trapping effects with kinetic transport phenomena.

To support the interpretation of such experiments, a range of diffusion-trapping models has been developed. One-dimensional diffusion-trapping frameworks can reproduce TDS spectra with reasonable accuracy but do not explicitly resolve microstructural features such as GBs, grain size, or defect connectivity \cite{GarcaMacas2024}. Phase-field methods offer a powerful alternative by enabling spatially resolved representations of microstructures and their influence on hydrogen transport \cite{Zhang2021, Hussein2024, Hussein2024b, Hussein2025}. However, conventional phase-field formulations rely on diffuse interfaces whose numerical thickness greatly exceeds the physical nanometer-scale thickness of GBs \cite{Feyen2023}. This may lead to artificial dependence of predicted hydrogen concentrations, fluxes, and kinetics on numerical interface thickness rather than on physical material parameters \cite{Hussein2024, Hussein2024b}. Such interface-scaling artifacts pose a fundamental obstacle to the quantitative interpretation of TDS and permeation experiments, and to reliable upscaling toward engineering-relevant length scales. Hussein et al.~recognized this issue and introduced an empirical scaling factor to mitigate the effect of the enlarged diffuse-interface width~\cite{Hussein2024b}. However, a thermodynamically consistent scaling formulation is still needed, together with benchmark tests demonstrating that the scaling preserves both the equilibrium GB hydrogen excess and the diffusion fluxes parallel and perpendicular to the GB.

The central problem addressed in this work is therefore the development of a \textit{quantitative} phase-field formulation for hydrogen diffusion and GB trapping, where ``quantitative'' means that total trapped hydrogen content in the GB and fluxes parallel and perpendicular to the GB are invariant with respect to the numerical thickness of the diffuse GB. This is essential if physical GBs with nanometer-scale thickness are to be represented on computational grids suitable for mesoscale polycrystalline microstructures. Within this framework, two application-oriented questions are investigated. First, how does GB diffusion influence the interpretation of TDS spectra and the apparent trapping energies obtained from Kissinger-type analyses? Second, how should effective hydrogen diffusion coefficients derived from permeation experiments be interpreted when GBs form an interconnected transport network rather than acting as isolated traps?

To address these questions, we develop a quantitative phase-field model for hydrogen diffusion and GB trapping that (i) enforces local thermodynamic equilibrium between bulk lattice and GB regions, (ii) preserves the physical GB volume (i.e., number of trapping sites) independent of numerical interface scaling, and (iii) accounts for anisotropic hydrogen mobility parallel and perpendicular to GBs. The model is formulated in a general thermodynamic and kinetic framework compatible with CALPHAD-based descriptions \cite{Lukas2007}, enabling application to a wide range of alloying systems. The microstructure is assumed to be static, and mechanical effects, such as stress-assisted diffusion and evolving defects, are intentionally excluded in order to isolate the influence of GBs on hydrogen transport.

The framework is systematically verified through benchmark problems designed to test equilibrium trapping, steady-state fluxes, transient kinetics, and numerical upscaling behavior in one, two, and three dimensions. It is then applied to two experimentally relevant problems: the interpretation of TDS spectra in the presence of GB diffusion, and the analysis of effective diffusion coefficients in polycrystalline microstructures as a function of grain size, trapping energy, temperature, and GB mobility. By addressing these problems within a single quantitative and microstructure-resolved framework, this work aims to bridge the gap between nanoscale trapping physics, mesoscale microstructural transport, and experimentally accessible observables relevant to hydrogen energy applications.

Readers primarily interested in the experimental implications of the model may proceed directly to Applications~I and~II. The model framework and benchmark sections provide the methodological basis and numerical verification of the approach, but the main physical findings concerning TDS interpretation and effective hydrogen diffusion are discussed in the application sections.

\begin{table*}[t]
\centering
\caption{List of symbols used in the model framework.}
\label{tab:symbols_model}
\begin{tabularx}{\textwidth}{@{}p{0.20\textwidth}p{0.15\textwidth}X@{}}
\toprule
\textbf{Symbol} & \textbf{Unit} & \textbf{Description} \\
\midrule

$\phi_i$ & $-$ & Non-conserved order parameter representing grain or phase $i$ \\
$N_\phi$ & $-$ & Number of order parameters \\
$h_\mathrm{B}$ & $-$ & Interpolation function representing the local bulk volume fraction \\
$h_\mathrm{GB}$ & $-$ & Interpolation function representing the local grain boundary volume fraction \\
$\tilde{\lambda}$ & $-$ & Normalized interfacial thickness \\
$t_\mathrm{int,PF}$ & $\mathrm{m}$ & Numerically selected phase-field interfacial thickness \\
$t_\mathrm{int,phys}$ & $\mathrm{m}$ & Physical grain boundary thickness \\
$C_\mathrm{int}$ & $-$ & Interfacial scaling factor used in the interpolation functions \\
$V_\mathrm{GB}$ & $\mathrm{m^3}$ & Total grain boundary volume \\
$A_\mathrm{GB}$ & $\mathrm{m^2}$ & Total grain boundary area \\
$V$ & $\mathrm{m^3}$ & Total system volume \\
$C_\mathrm{H}$ & $\mathrm{mol\,m^{-3}}$ & Total hydrogen concentration \\
$C_\mathrm{H,B}$ & $\mathrm{mol\,m^{-3}}$ & Hydrogen concentration in the bulk lattice \\
$C_\mathrm{H,GB}$ & $\mathrm{mol\,m^{-3}}$ & Hydrogen concentration in the grain boundary \\
$y_\mathrm{H}$ & $-$ & Total lattice site fraction of hydrogen \\
$y_\mathrm{H,B}$ & $-$ & Lattice site fraction of hydrogen in the bulk \\
$y_\mathrm{H,GB}$ & $-$ & Lattice site fraction of hydrogen in the grain boundary \\
$f_\mathrm{lattice}$ & $-$ & Sublattice fraction, equal to $n/m$ \\
$m$ & $-$ & Number of host-lattice atoms per unit cell \\
$n$ & $-$ & Number of interstitial sites per unit cell \\
$V_\mathrm{UC}$ & $\mathrm{m^3}$ & Unit-cell volume \\
$V_\mathrm{m}$ & $\mathrm{m^3\,mol^{-1}}$ & Molar volume of the alloy in the absence of hydrogen \\
$N_\mathrm{A}$ & $\mathrm{mol^{-1}}$ & Avogadro constant \\
$N_\mathrm{T}$ & $\mathrm{mol\,m^{-3}}$ & Total amount of grain boundary trapping sites per unit volume \\
$\mu_\mathrm{H}$ & $\mathrm{J\,mol^{-1}}$ & Chemical potential of hydrogen \\
$\mu_\mathrm{H,B}$ & $\mathrm{J\,mol^{-1}}$ & Chemical potential of hydrogen in the bulk \\
$\mu_\mathrm{H,GB}$ & $\mathrm{J\,mol^{-1}}$ & Chemical potential of hydrogen in the grain boundary \\
$\mu_\mathrm{H,G}$ & $\mathrm{J\,mol^{-1}}$ & Chemical potential of hydrogen in the gas phase \\
$\mu_\mathrm{H,B,0}$ & $\mathrm{J\,mol^{-1}}$ & Reference contribution to the bulk hydrogen chemical potential \\
$\Delta\mu_\mathrm{H,b,GB}$ & $\mathrm{J\,mol^{-1}}$ & Grain boundary hydrogen binding free energy \\
$E_\mathrm{trap}$ & $\mathrm{J\,mol^{-1}}$ & Trap binding-energy magnitude \\
$R$ & $\mathrm{J\,mol^{-1}\,K^{-1}}$ & Gas constant \\
$T$ & $\mathrm{K}$ & Temperature \\
$M_\mathrm{at,H}$ & $\mathrm{m^2\,mol\,J^{-1}\,s^{-1}}$ & Atomic mobility of hydrogen \\
$\mathbf{M}_\mathrm{PF,H}$ & $\mathrm{m^2\,mol\,J^{-1}\,s^{-1}}$ & Phase-field hydrogen mobility tensor \\
$M_\perp$ & $\mathrm{m^2\,mol\,J^{-1}\,s^{-1}}$ & Mobility component perpendicular to the grain boundary \\
$M_\parallel$ & $\mathrm{m^2\,mol\,J^{-1}\,s^{-1}}$ & Mobility component parallel to the grain boundary \\
$M_\mathrm{at,H,B}$ & $\mathrm{m^2\,mol\,J^{-1}\,s^{-1}}$ & Atomic mobility of hydrogen in the bulk \\
$M_\mathrm{at,H,GB}$ & $\mathrm{m^2\,mol\,J^{-1}\,s^{-1}}$ & Atomic mobility of hydrogen in the grain boundary \\
$M_\mathrm{at,H,B,0}$ & $\mathrm{m^2\,mol\,J^{-1}\,s^{-1}}$ & Pre-factor contribution to the bulk atomic mobility of hydrogen \\
$M_\mathrm{at,H,GB,0}$ & $\mathrm{m^2\,mol\,J^{-1}\,s^{-1}}$ & Pre-factor contribution to the grain boundary atomic mobility of hydrogen \\
$\mathbf{n}_\mathrm{GB}$ & $-$ & Unit normal vector to the grain boundary \\
$\mathbf{I}$ & $-$ & Identity tensor \\
$J_\perp$ & $\mathrm{mol\,m^{-2}\,s^{-1}}$ & Hydrogen flux density perpendicular to the grain boundary \\
$J_\parallel$ & $\mathrm{mol\,m^{-2}\,s^{-1}}$ & Hydrogen flux density parallel to the grain boundary \\
$t$ & $\mathrm{s}$ & Time \\

\bottomrule
\end{tabularx}
\end{table*}

\begin{table*}[t]
\centering
\caption{List of symbols used in the benchmarks and applications.}
\label{tab:symbols_applications}
\begin{tabularx}{\textwidth}{@{}p{0.20\textwidth}p{0.15\textwidth}X@{}}
\toprule
\textbf{Symbol} & \textbf{Unit} & \textbf{Description} \\
\midrule

$d_\mathrm{grain}$ & $\mathrm{m}$ & Average grain size \\
$L_x,L_y,L_z$ & $\mathrm{m}$ & Sample dimensions in the $x$-, $y$-, and $z$-directions \\
$\Delta x$ & $\mathrm{m}$ & Grid spacing \\
$n_x,n_y,n_z$ & $-$ & Number of grid points in the $x$-, $y$-, and $z$-directions \\
$C_{\mathrm{H,GB,eq}}$ & $\mathrm{mol\,m^{-3}}$ & Equilibrium hydrogen concentration in the grain boundary \\
$N_\mathrm{H}$ & $\mathrm{mol}$ & Total amount of hydrogen in the sample \\
$N_\mathrm{H,GB}$ & $\mathrm{mol}$ & Amount of hydrogen trapped at grain boundaries \\
$r_\mathrm{M,GB}$ & $-$ & Ratio between grain boundary and bulk hydrogen mobility \\
$D_\mathrm{B,0}$ & $\mathrm{m^2\,s^{-1}}$ & Bulk hydrogen diffusion prefactor  \\
$E_\mathrm{migr}$ & $\mathrm{J\,mol^{-1}}$ & Migration energy for lattice diffusion of hydrogen \\
$D_\mathrm{H,B}$ & $\mathrm{m^2\,s^{-1}}$ & Chemical diffusion coefficient of hydrogen in the bulk \\
$D_\mathrm{H,GB}$ & $\mathrm{m^2\,s^{-1}}$ & Chemical diffusion coefficient of hydrogen in the grain boundary \\
$D_\mathrm{eff,ss}$ & $\mathrm{m^2\,s^{-1}}$ & Effective steady-state diffusion coefficient \\
$D_\mathrm{eff,ts}$ & $\mathrm{m^2\,s^{-1}}$ & Effective transient-state diffusion coefficient \\
$D_\mathrm{eff,Or}$ & $\mathrm{m^2\,s^{-1}}$ & Oriani effective diffusion coefficient \\
$J_\mathrm{H}$ & $\mathrm{mol\,m^{-2}\,s^{-1}}$ & Hydrogen flux density \\
$J_\mathrm{H,\perp}$ & $\mathrm{mol\,m^{-2}\,s^{-1}}$ & Hydrogen flux density perpendicular to the grain boundary \\
$J_\mathrm{H,\parallel}$ & $\mathrm{mol\,m^{-2}\,s^{-1}}$ & Hydrogen flux density parallel to the grain boundary \\
$J_\mathrm{H,B}$ & $\mathrm{mol\,m^{-2}\,s^{-1}}$ & Hydrogen flux density through the bulk \\
$J_\mathrm{H,GB}$ & $\mathrm{mol\,m^{-2}\,s^{-1}}$ & Hydrogen flux density through the grain boundary \\
$J_\mathrm{H,\parallel,total}$ & $\mathrm{mol\,s^{-1}}$ & Total hydrogen flux parallel to the grain boundary \\
$J_\mathrm{H,total,ss}$ & $\mathrm{mol\,s^{-1}}$ & Steady-state total hydrogen flux \\
$t_\mathrm{lag}$ & $\mathrm{s}$ & Lag time used to determine the transient effective diffusion coefficient \\
$P_\mathrm{H}$ & $\mathrm{Pa}$ & Hydrogen partial pressure \\
$P_\mathrm{Char}$ & $\mathrm{Pa}$ & Hydrogen pressure during charging \\
$P_\mathrm{UHV}$ & $\mathrm{Pa}$ & Hydrogen pressure under ultra-high vacuum desorption conditions \\
$\dot{T}$ & $\mathrm{K\,s^{-1}}$ & Heating rate \\
$T_\mathrm{peak}$ & $\mathrm{K}$ & TDS peak temperature \\
$J_\mathrm{H,peak}$ & $\mathrm{mol\,s^{-1}}$ & Maximum hydrogen desorption flux in a TDS spectrum \\
$J_\mathrm{TDS,eq}$ & $\mathrm{mol\,s^{-1}}$ & Equilibrium TDS desorption flux \\
$E_\mathrm{a}$ & $\mathrm{J\,mol^{-1}}$ & Apparent activation energy obtained from Kissinger analysis \\
$K_\mathrm{GB}$ & $-$ & Grain boundary trapping equilibrium constant \\
$\theta_\mathrm{T}$ & $-$ & Trap occupancy \\
$c_\mathrm{T}$ & $\mathrm{mol\,m^{-3}}$ & Average trapped hydrogen concentration \\
$c_\mathrm{L}$ & $\mathrm{mol\,m^{-3}}$ & Average lattice-bound hydrogen concentration \\

\bottomrule
\end{tabularx}
\end{table*}

\section{Model framework}\label{sec:Model}
In this section, we describe the proposed phase-field model used for hydrogen diffusion and trapping. Different grains or phases of the material are described by non-conserved order parameters $\phi_i$. For most phase-field models, these fields evolve in time to describe phase transformations or grain growth by means of the Allen-Cahn equation \cite{Allen1972, Moelans2008-intro}. For diffusion and trapping of hydrogen, however, we neglect the movement of these grains/phases, because the low temperatures considered here make GBs immobile. The order parameter profiles are still necessary to describe the GBs, and thus the microstructure. Therefore, the microstructure is generated using a different phase-field model, previously developed within our group \cite{Verbeeck2024}. By controlling the input parameters of this model, we can obtain a realistic equiaxed microstructure with a selected mean grain size. 

The physical GBs of interest have characteristic thicknesses on the nanometer scale, whereas phase‑field simulations typically resolve microstructures at much larger length scales. In this context, a parameter called the normalized interfacial thickness $\tilde{\lambda}$\footnote{The symbols used in this work are summarized in Tables~\ref{tab:symbols_model} and~\ref{tab:symbols_applications}.} as defined by Feyen et al. \cite{Feyen2023}, will be used to describe the scaling of the simulation:
\begin{equation}
\label{eq:normalized_interfacial_thickness}
\begin{split}
\tilde{\lambda} = \frac{t_\mathrm{int,PF}}{t_\mathrm{int,phys}}.\\
\end{split}
\end{equation}
Here, the $t_\mathrm{int,PF}$ is the numerically selected interfacial thickness of the order parameter profile, which can range from millimeters to nanometers depending on the grain size,  and $t_\mathrm{int,phys}$ is the true grain/phase boundary thickness, which is on the order of nanometers \cite{Porter2009}. 

\subsection{Physical assumptions and modeling choices}
The phase-field model is formulated under the following physical assumptions, which define its range of applicability to hydrogen diffusion, trapping, and desorption in polycrystalline metals:

%\begin{itemize}

\textbf{Static microstructure:} The microstructure is assumed to remain stationary during hydrogen diffusion and thermal desorption. GB motion, grain growth, and phase transformations are neglected. Phase-field order parameters are therefore used solely to represent the spatial distribution of grains and GBs.

\textbf{Grain boundary thermodynamics:} GBs are treated as regions with a finite physical thickness (on the order of nanometers~\cite{Hamza2015}) and thermodynamic properties distinct from the bulk lattice. Hydrogen experiences an altered chemical potential in the GB, represented through a binding free energy. The total physical GB volume, and thus the number of trapping sites, is assumed to be fixed and independent of numerical interface scaling.

\textbf{Interstitial diffusion:} Hydrogen diffuses exclusively as an interstitial species in a stationary host lattice. Substitutional transport and lattice motion are neglected. The molar volume is assumed to be independent of hydrogen concentration, which is justified by the low solubility of hydrogen in endothermic occluders such as iron-based alloys \cite{Cotterill1961}.

\textbf{Local thermodynamic equilibrium:} Hydrogen occupying bulk lattice sites and GB sites is assumed to be in local thermodynamic equilibrium at every spatial location, implying equality of chemical potentials. This corresponds to Oriani's local equilibrium condition \cite{Oriani1970} and is justified when trapping and detrapping kinetics are fast in relation to diffusion. Such conditions are expected for nanometer-scale GB regions, where atomic jump frequencies are sufficiently high for local equilibrium to be rapidly established, as supported by recent analyses based on the McNabb-Foster model \cite{Mancias2022}.

\textbf{Anisotropic mobility:} Hydrogen mobility is allowed to differ parallel and perpendicular to GBs. Transport perpendicular to the GB is assumed to be governed by bulk diffusion, neglecting any additional interface influence due to the small physical thickness of the GB relative to the grid spacing. GB-assisted diffusion is therefore confined to directions parallel to the boundary plane and can be larger or smaller compared to the bulk diffusivity.

\textbf{Neglect of mechanical effects:} Stress-assisted diffusion, elastic contributions to the hydrogen chemical potential, and hydrogen-stress coupling are neglected. The system is assumed to be mechanically unloaded.

\textbf{Single trap type:} GBs are the only explicitly modeled trapping sites. Other defects, e.g., dislocations, vacancies, and precipitates, are not considered here, but could be incorporated in future extensions of the model. The number of trapping sites is assumed to remain constant during the simulation. As a first step, only one general GB trapping free energy is considered here, whereas in reality, the properties for each GB type differ significantly \cite{Ding2019, McEniry2017, Hussein2024b}.

\textbf{Thermal conditions:} Temperature is assumed to be spatially uniform and varies only as a function of time according to a prescribed heating rate, consistent with TDS and permeation cell conditions.

\textbf{Boundary conditions:} At external surfaces, hydrogen is assumed to be in local thermodynamic equilibrium with the surrounding gas phase, described by a prescribed chemical potential of hydrogen. This boundary condition governs hydrogen exchange during charging and desorption, enabling direct representation of experimental conditions such as thermal desorption spectroscopy and permeation experiments.

Under these assumptions, the model is applicable to hydrogen diffusion and trapping in single-phase, polycrystalline metals with static microstructures. It is particularly suited for single-phase polycrystalline metals in which GBs act as reversible trapping sites and interconnected transport pathways, such as ferritic and austenitic steels, and single-phase nickel alloys.

\subsection{Grain boundary trapping and thermodynamics}
Hydrogen will diffuse interstitially in most metals \cite{Yu2024}. This can be modeled by a sublattice model, as defined by the CALPHAD method \cite{Lukas2007}: $(M_1,M_2,...)_m:(H,I_2,I_3,...,\mathrm{Va})_n$. With $M_1$, $M_2$, ... the metal atoms in the sublattice 1, $H$ the hydrogen atoms, $I_2,I_3,...$ other interstitial atoms, $\mathrm{Va}$ the vacancies in the sublattice 2 (i.e., empty interstitial sites), $m$ the number of atoms in sublattice 1 in the unit cell, and $n$ the number of interstitial sites in the unit cell (BCC m=2, tetrahedral sublattice: n=12 , octahedral sublattice: n=6 and for FCC m=4, tetrahedral sublattice: n=8 , octahedral sublattice: n=4): . In this case, we assume that the first sublattice is always fully occupied (i.e., there is a negligible vacancy concentration). Hydrogen can only occupy the interstitial sublattice 2. Note that additional atoms, such as carbon, can be present in sublattice 2 as well. We define the concentration of hydrogen $C_\mathrm{H}$, using the sublattice description as:
\begin{equation}
\label{eq:C_H}
\begin{split}
C_\mathrm{H} &=  \frac{n y_\mathrm{H}}{V_\mathrm{UC}N_\mathrm{A}}\\
&= \frac{n y_\mathrm{H}}{mV_\mathrm{m}}\\
&= f_\mathrm{lattice}\frac{ y_\mathrm{H}}{V_\mathrm{m}}.\\
\end{split}
\end{equation}
Here, $y_\mathrm{H}$ is the site fraction of hydrogen in sublattice 2, $V_\mathrm{m}$ the molar volume of the alloy in the absence of hydrogen, $V_\mathrm{UC}$ the volume of the unit cell, and the sublattice fraction $f_\mathrm{lattice} = n/m$ which is a constant depending on the sublattice model. It is assumed that the molar volume $V_\mathrm{m}$ does not change significantly with the addition of hydrogen, which is justified when the alloy exhibits a low solubility of hydrogen. 

In the proposed phase-field model, we apply the Kim-Kim-Suzuki (KKS) model, widely known to be used for describing diffusion in the interface of two phases during diffusion-controlled phase transformation in a quantitative manner \cite{Kim1999}. In the proposed model, we solely apply the KKS model to the GB, treating it effectively as a different phase. We justify this by stating that within the GB structure, the chemical potential of hydrogen is different from the bulk; hence it acts as a trapping site when it is lower compared to the bulk. The total hydrogen concentration ($C_\mathrm{H}$) can then be split into the bulk concentration ($C_\mathrm{H,B}$) and the GB concentration ($C_\mathrm{H,GB}$). We can define the first auxiliary equation of the KKS model, which describes the mass balance as: 
\begin{equation}
\label{eq:mass_balance}
\begin{split}
C_\mathrm{H} &=  h_\mathrm{B} C_\mathrm{H,B} +  h_\mathrm{GB} C_\mathrm{H,GB}. \\
\end{split}
\end{equation}
Equivalently, using Equation~\eqref{eq:C_H}:
\begin{equation}
\begin{split}
y_\mathrm{H} &=  h_\mathrm{B} y_\mathrm{H,B} +  h_\mathrm{GB} y_\mathrm{H,GB}. \\
\end{split}
\end{equation}
In this equation $y_\mathrm{H,B}$ and $y_\mathrm{H,GB}$ represent the lattice site fraction of hydrogen in the bulk and the GB, respectively. $h_\mathrm{B}$ and $h_\mathrm{GB}$ are specifically chosen interpolation functions in function of the order parameters $\phi_{i,j,...}$, representing the local volume fraction of the bulk phase and the GB, such that: 
\begin{equation}
\begin{split}
h_\mathrm{B} +  h_\mathrm{GB}= 1. \\
\end{split}
\end{equation}
In this work, interpolation functions were defined in the style proposed by Moelans, often used to represent the volume fraction of bulk phases \cite{Moelans2011}. Here, these functions are extended to treat GBs as a separate phase-like contribution, which requires interaction terms between pairs of order parameters. In addition, an interfacial scaling factor is introduced to preserve the physical GB volume when the phase-field interface thickness is scaled:
\begin{equation}
\label{eq:h_B}
\begin{split}
h_\mathrm{B} &= \frac{\sum^{N_\phi}_{i = 1} \phi_i^2}{\sum^{N_\phi}_{i = 1} \left(\phi_i^2 +  \sum^{N_\phi}_{j \neq i}\frac{C_\mathrm{int}(\tilde{\lambda})}{\tilde{\lambda}}\phi_i \phi_j \sum^{N_\phi}_{k \neq i,j}\left(1 -  \phi_k\right)\right)},\\ 
\end{split}
\end{equation}
\begin{equation}
\label{eq:h_GB}
\begin{split}
h_\mathrm{GB} &= \frac{\sum^{N_\phi}_{i = 1} \sum^{N_\phi}_{j \neq i} \frac{C_\mathrm{int}(\tilde{\lambda})}{\tilde{\lambda}}\phi_i \phi_j\sum^{N_\phi}_{k \neq i,j}\left(1 -  \phi_k\right)}{\sum^{N_\phi}_{i = 1} \left(\phi_i^2 +  \sum^{N_\phi}_{j \neq i}\frac{C_\mathrm{int}(\tilde{\lambda})}{\tilde{\lambda}}\phi_i \phi_j \sum^{N_\phi}_{k \neq i,j}\left(1 -  \phi_k\right)\right)}.\\ 
\end{split}
\end{equation}
Here, \(C_\mathrm{int}\) is an interfacial scaling factor that depends only on the simulation scale through \(\tilde{\lambda}\). $C_\mathrm{int}$ is defined such that the true volume of  all grain boundaries in a simulation $V_\mathrm{GB}$ is independent of the scale of the phase-field interfacial thickness $t_\mathrm{int,PF}$ (assuming that the interface thickness is small compared with the local radius of curvature of the GB, \(t_\mathrm{int,PF}/R_\mathrm{GB}\ll 1\)):
\begin{equation}
\label{eq:V_GB}
\begin{split}
V_\mathrm{GB}  &= \int_V h_\mathrm{GB}(\tilde{\lambda}) \mathrm{d}x\mathrm{d}y\mathrm{d}z\\
&=A_\mathrm{GB}t_\mathrm{int,phys}\neq f(\tilde{\lambda}).\\
\end{split}
\end{equation}
Here, \(x\), \(y\), and \(z\) are the spatial coordinates of the system, and \(A_\mathrm{GB}\) is the total GB area. The values of $C_\mathrm{int}(\tilde{\lambda})$ are calculated numerically based on the chosen normalized interface thickness $\tilde{\lambda}$ of a system containing a single flat grain boundary (see Appendix~A).  The independence of the GB volume from the phase-field interfacial thickness is of utmost importance to provide quantitative results from the model and to make sure that the total amount of GB trapping sites per unit volume ($N_\mathrm{T}$) will not depend on the resolution of the simulation ($t_\mathrm{int,PF}$). The total amount of GB trapping sites can be calculated as follows:
\begin{equation}
\begin{split}
N_\mathrm{T}=f_\mathrm{lattice}\frac{V_\mathrm{GB}}{V}\frac{1}{V_\mathrm{m}}.
\end{split}
\end{equation}
Here, \(V\) is the total system volume.

Note that the interpolation functions also incorporate a correction term $\left(1 -  \phi_k\right)$, which reduces excess trapping sites at triple junctions resulting from overlapping order parameters.

The second auxiliary equation of the KKS model assumes local equilibrium, often described as the equality of diffusion potentials, between two or more phases for substitutional diffusion. In the case of interstitial diffusion, this translates into the equality between the chemical potentials of hydrogen in the bulk phase $\mu_\mathrm{H, B}$ and the GBs $\mu_\mathrm{H, GB}$:
\begin{equation}
\label{eq:KKS}
\begin{split}
\mu_\mathrm{H,B}(C_\mathrm{H,B}) &= \mu_\mathrm{H,GB}(C_\mathrm{H,GB})=\mu_\mathrm{H}. \\
\end{split}
\end{equation}
In which $\mu_\mathrm{H}$ is the chemical potential of hydrogen in the system, which drives diffusion. At every position in the microstructure, hydrogen will be distributed over the GB (where the GB fraction $h_\mathrm{GB}>0$) and the bulk phase, assuming local equilibrium (based on Equation~\eqref{eq:KKS} and \eqref{eq:mass_balance}), similar to Oriani's assumption \cite{Oriani1970}. Note that the proposed model can be extended straightforwardly to consider different types of GBs, each with a different chemical potential and mobility, by assigning separate interpolation functions to specific pairs of order parameters. In that case, Equation~\eqref{eq:mass_balance} and Equation~\eqref{eq:KKS} would contain additional terms.
 
The chemical potential of hydrogen in the bulk phase for a given alloy composition and temperature is modeled by a regular solution model. This model was identified as the formulation used in the Thermo-Calc Software TCFE13 database~\cite{TCFE13}. The bulk chemical potential of hydrogen is given by:
\begin{equation}
\label{eq:chempot_B}
\begin{split}
\mu_\mathrm{H,B} =& \mu_\mathrm{H,B,0}(T,y_\mathrm{M_1,B},y_\mathrm{M_2,B},...) +RT\ln{y_\mathrm{H,B}} \\
&- RT\ln{\left(1 - y_\mathrm{H,B} - y_\mathrm{I_2,B}-...\right)}.\\
\end{split}
\end{equation}
Here, $R$ is the universal gas constant, T is the temperature,  $y_\mathrm{M_1,B}$, $y_\mathrm{M_2,B}$, ...  are the lattice site fraction of the metal atoms in sublattice 1 in the bulk and $y_{I_2,B}$, ... are the interstitial atoms in sublattice 2 in the bulk phase. The reference bulk chemical-potential parameter $\mu_{\mathrm{H,B,0}}$ can be extracted from thermodynamic databases, such as TCFE13~\cite{TCFE13}, as done in this work.

The chemical potential of H in the GB is chosen to be modeled similarly to the bulk phase, with the addition of the hydrogen binding free energy of the GB. The binding free energy is written as \(\Delta\mu_\mathrm{H,b,GB}=\mu_\mathrm{H,GB}-\mu_\mathrm{H,B}\), with attractive trapping corresponding to \(\Delta\mu_\mathrm{H,b,GB}<0\). The commonly reported trapping energy magnitude is then \(E_\mathrm{trap}=-\Delta\mu_\mathrm{H,b,GB}\) when entropic contributions are neglected. However, any description for $\Delta\mu_\mathrm{H,b,GB}$ can be used, which can depend, for example, on temperature, as recently shown by MD simulations of the iron-hydrogen system, in which the trap binding free energy was calculated \cite{Starikov2022}. Using the trap binding free energy, the chemical potential of H in the GB can be defined as:
\begin{equation}
\label{eq:chempot_GB}
\begin{split}
\mu_\mathrm{H,GB} =& \mu_\mathrm{H,B,0}(T,y_\mathrm{M_1,GB},y_\mathrm{M_2,GB},...) +RT\ln{y_\mathrm{H,GB}} \\
&- RT\ln{\left(1 - y_\mathrm{H,GB} - y_\mathrm{I_2,GB} - ...\right)} + \Delta\mu_\mathrm{H,b,GB}.\\
\end{split}
\end{equation}
With $y_\mathrm{M_1,GB}$, $y_\mathrm{M_2,GB}$, ... the lattice site fraction of the metal atoms in sublattice 1 in the GB and $y_\mathrm{I_2,GB}$, ... the interstitial atoms in sublattice 2 in the GB. Note that lattice site fractions are often referred to as occupancies ($\theta_\mathrm{T}$ for a trap and $\theta_\mathrm{L}$ for the lattice) in literature concerning hydrogen \cite{Yu2024}.

\subsection{Diffusion formulation and kinetics}
\begin{figure}[pos=h]
	\centering
		\includegraphics[width=0.8\linewidth]{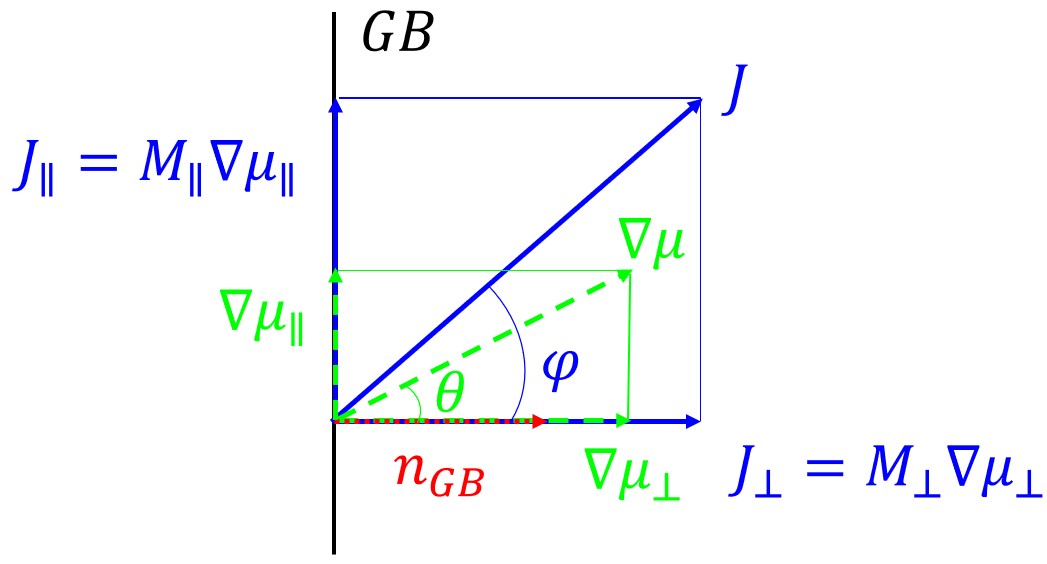}
	\caption{Projection of the mobility onto the GB, which results in a decomposition of the overall flux into the perpendicular flux   $J_\perp$ and the parallel flux $J_\parallel$.}
	\label{FIG:Illustration_J}
\end{figure}

As discussed in the previous paragraph the model is based on two sublattices: a host lattice of metallic atoms, assumed to be static and an interstitial sublattice containing species like hydrogen. Using the diffusion equation for interstitial transport in the lattice‑fixed frame, with the host lattice taken as stationary, we derive, following Andersson and Ågren \cite{Andersson1992}:
\begin{equation}
\begin{split}
\frac{\partial C_\mathrm{H}}{\partial t} &= \nabla \cdot \left(M_\mathrm{at,H}\: C_\mathrm{H} \nabla \mu_\mathrm{H}\right).\\
\end{split}
\end{equation}
Here, $M_\mathrm{at,H}$ denotes the atomic mobility of hydrogen. In the phase-field formulation, however, we introduce a phase-field mobility  $\mathbf{M}_\mathrm{PF,H}$, which is a tensor, to account for anisotropic hydrogen diffusion in the GBs, with different mobilities parallel and perpendicular to the GB plane. In addition, spatial variations in both atomic mobility and hydrogen concentration between the bulk and the GB are included in the phase-field mobility. To incorporate the anisotropy effect, a projection of the mobility tensor onto the local GB normal is applied, as schematically illustrated in Figure~\ref{FIG:Illustration_J}. Similar projection approaches were also used by other authors \cite{Lvov2022, Ahmed2015}. 
Expressing the hydrogen concentration in terms of the lattice site fraction using Equation~\eqref{eq:C_H}, we can obtain the final diffusion equation:
\begin{equation}
\label{eq:pf_diffusion}
\begin{split}
\frac{\partial y_\mathrm{H}}{\partial t} &= \nabla \cdot \left(\mathbf{M}_\mathrm{PF,H}\nabla \mu_\mathrm{H}\right)\\
\end{split}
\end{equation}
With $\mathbf{M}_\mathrm{PF,H}$:
\begin{equation}
\begin{split}
\mathbf{M}_\mathrm{PF,H} =&  M_{\perp}\left(\mathbf{n}_\mathrm{GB} \otimes  \mathbf{n}_\mathrm{GB} \right) +  M_{\parallel}(\mathbf{I} - \mathbf{n}_\mathrm{GB} \otimes \mathbf{n}_\mathrm{GB})\\
\end{split}
\end{equation}
Here $\otimes$ denotes the tensor (outer) product, and $\mathbf{n}_\mathrm{GB}$ is the unit normal vector to the GB. The perpendicular and parallel mobility components $M_{\perp}$ and $M_\parallel$ are defined as: 
\begin{equation}
\label{eq:M_perp}
\begin{split}
M_{\perp} =& M_\mathrm{at,H,B}y_\mathrm{H,B},\\
\end{split}
\end{equation}
and
\begin{equation}
\label{eq:M_para}
\begin{split}
M_{\parallel} &=  h_\mathrm{B} M_\mathrm{at,H,B}y_\mathrm{H,B} + h_\mathrm{GB} M_\mathrm{at,H,GB}y_\mathrm{H,GB}.\\
\end{split}
\end{equation}
This formulation ensures that diffusion perpendicular to the GB is governed solely by the bulk atomic mobility, independent of the chosen phase-field interfacial thickness. It is thus assumed that diffusion through the nanometer-sized physical GB would be of negligible influence on the perpendicular diffusion. This assumption is supported by the extreme aspect ratio between the GB thickness ($\approx$ 1 nm \cite{Hamza2015}) and diffusion distances in TDS experiments ($\SI{}{\micro m}$ - mm). Nevertheless, the proposed framework is sufficiently flexible to accommodate alternative physical assumptions. For instance, if the detrapping barrier is considered to be very high, the perpendicular flux can be further reduced by incorporating this effect directly into Equation~\eqref{eq:M_perp}. The diffusion parallel to the GB is controlled by a combination of bulk and GB diffusion, while remaining independent of the numerical value of the phase-field interface thickness  $t_\mathrm{int,PF}$, due to the nature of the proposed interpolation functions $h_\mathrm{B}$ and $h_\mathrm{GB}$.

The atomic mobility, according to the definition of Andersson and Ågren, contains the vacant lattice site fraction \cite{Andersson1992}. In our case, we model this explicitly, due to the fact that for a high hydrogen binding free energy ($\Delta\mu_\mathrm{H,b,GB}$), the trapping site will be nearly full, and the number of empty interstitial sites will be limiting the diffusion through the GB:
\begin{equation}
\begin{split}
M_\mathrm{at,H,B} =& M_\mathrm{at,H,B,0}(T,y_\mathrm{M_{1,B}},y_\mathrm{M_{2,B}},...)(1-y_\mathrm{H,B}) \\
\end{split}
\end{equation}
and in the GB:
\begin{equation}
\begin{split}
M_\mathrm{at,H,GB} =& M_\mathrm{at,H,GB,0}(T,y_\mathrm{M_{1,GB}},y_\mathrm{M_{2,GB}},...)(1-y_\mathrm{H,GB}) \\
\end{split}
\end{equation}
The parameter \(M_{\mathrm{at,H,B,0}}\) can be obtained directly from kinetic CALPHAD databases, whereas \(M_{\mathrm{at,H,GB,0}}\) must be estimated from atomistic methods, such as molecular dynamics or DFT calculations~\cite{Starikov2022, Hamza2015}. In this work, the effect of this parameter is investigated systematically. Note that the atomic mobility is related to the tracer diffusion coefficient through the Einstein relation, \(M_\mathrm{at,H,B,0} = D^*_\mathrm{H,B}/(RT)\), when the driving force is expressed as a gradient in chemical potential~\cite{Mehrer2007}. In the dilute limit, where the thermodynamic factor approaches unity, the tracer diffusion coefficient is equal to the chemical diffusion coefficient. This approximation is appropriate for hydrogen diffusion in the bulk considered here. 

\subsection{Numerical implementation}
The governing equations are solved using an explicit finite‑difference scheme on a regular Cartesian grid. At each time step, the total hydrogen site fraction is advanced using the diffusion equation, after which local redistribution between bulk lattice and GB sites is enforced under the assumption of local thermodynamic equilibrium. The detailed derivation of the update equations and the numerical solution strategy are provided in Appendix~B.

To enable efficient simulations on realistic three- \\ dimensional microstructures, the model was implemented in a parallelized framework targeting graphics processing units (GPUs). The core numerical routines were developed using the CUDA framework (NVIDIA) and compiled into a MEX executable, allowing simulation setup, execution, visualization, and post‑processing to be performed within the MATLAB environment. In addition, a CPU‑based MATLAB implementation was developed and used for verification and debugging purposes. All scripts are freely available at Mendeley Data \cite{feyen2026_MendeleyData}.

The numerical framework supports a range of boundary conditions to accommodate different experimental and modeling scenarios. Periodic boundary conditions, Neumann boundary conditions with zero diffusive flux, and Dirichlet boundary conditions applied to the bulk hydrogen lattice site fraction can be specified by the user. In addition, local thermodynamic equilibrium between the bulk lattice and the external hydrogen gas can be assumed at these boundaries through a prescribed expression for the chemical potential of hydrogen in the gas phase, $\mu_\mathrm{H,G}$, as a function of temperature and hydrogen partial pressure. This flexibility enables direct simulation of hydrogen charging, permeation, and thermal desorption spectroscopy (TDS) experiments.

\section{Model verification and benchmarks}\label{sec:Model_validation}
To verify the proposed framework for quantitative GB diffusion, we introduce a set of benchmark problems designed to test the quantitative behavior of the model. Experimental validation was not pursued in this work for several reasons. First, such validation would require carefully controlled TDS and permeation experiments on well-characterized, pristine samples. Second, thermodynamic and kinetic GB parameters are required as model inputs with sufficient accuracy. Third, the present model explicitly considers only GBs as trapping sites, while other defects, such as dislocations, vacancies, and precipitates, are not included. Finally, as will be shown in Application~I, room-temperature TDS measurements may not reveal a distinct GB-related signal if GB diffusion contributes significantly to hydrogen transport. For these reasons, the model is verified using benchmark problems that isolate the relevant physical mechanisms and allow direct comparison with analytical solutions.
In the following benchmarks and applications, the GB mobility is treated as an unknown parameter. For simplicity, it is expressed relative to the bulk hydrogen mobility. We therefore define the GB mobility ratio as $ r_\mathrm{M,GB} = M_\mathrm{at,H,GB,0}/M_\mathrm{at,H,B,0}$, which is used as a control parameter in the parametric studies and benchmarks. Although \(M_\mathrm{at,H,GB,0}\) is generally not known a priori, the model remains useful because it enables its effect to be explored systematically at experimentally relevant microstructural length scales. The following benchmarks demonstrate that the framework can be applied directly at the experimental or microstructural scale, while Applications~I and~II show how it can support the interpretation of TDS and permeation experiments.

\subsection{Benchmark 1: Equilibrium hydrogen trapping at the grain boundary}
\begin{figure}[pos=h]
	\centering
		\includegraphics[width=\linewidth]{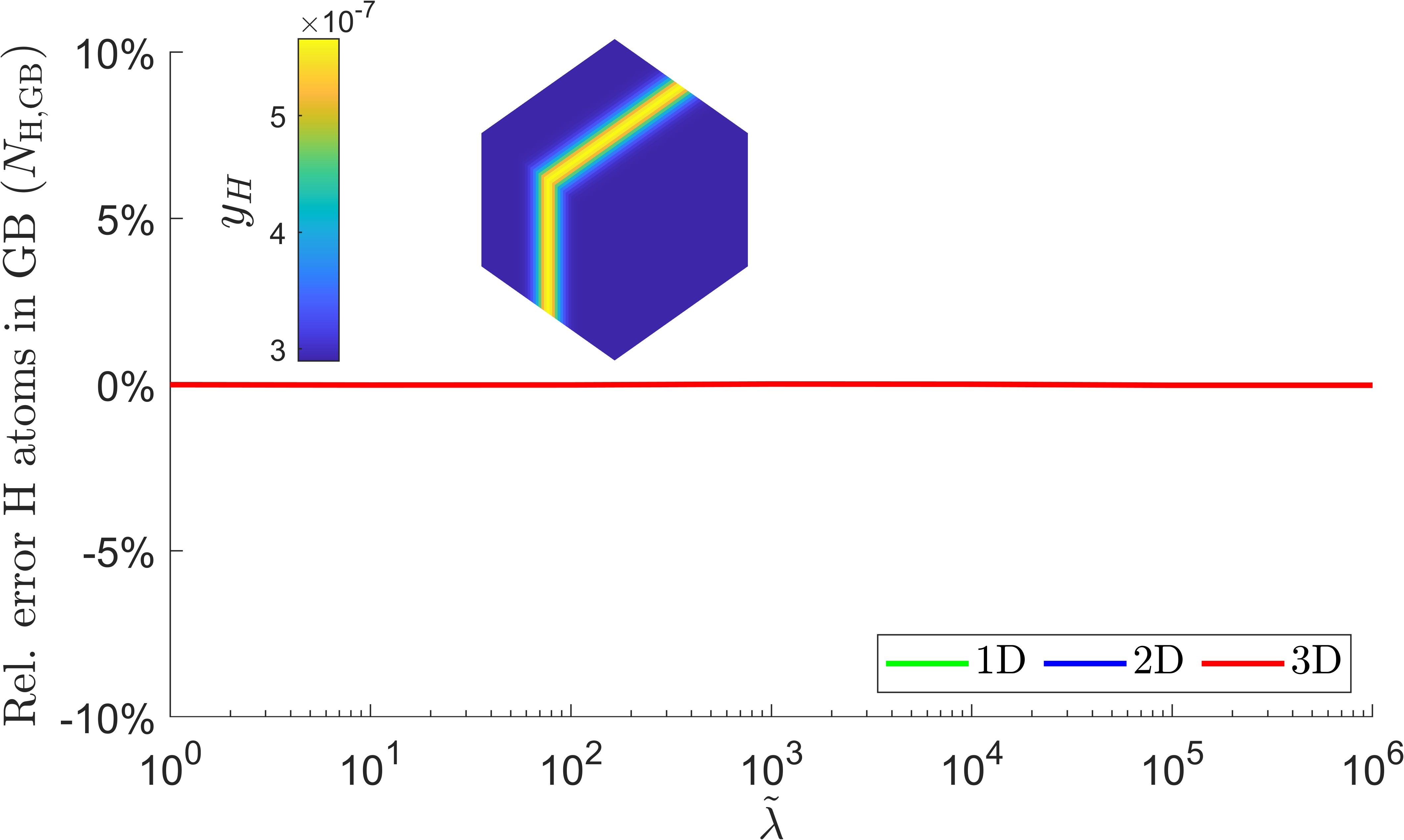}
	\caption{Benchmark 1: Relative error in the total amount of hydrogen trapped at the GB compared to the analytical solution as a function of the normalized interfacial thickness. The error remains below 0.02\% for all cases, indicating that all formulations are equivalent for practical applications.}
	\label{FIG:BM1}
\end{figure}

The first benchmark is intentionally easy to interpret. A domain is constructed containing a single GB located at the center of the system, represented by the overlap of two order parameters (see Figure~\ref{FIG:BM1}). A constant hydrogen concentration in the bulk ($C_\mathrm{H,B}$) is applied at the external boundaries of the domain, and the system is allowed to equilibrate. During equilibration, hydrogen partitions between the bulk and the GB until global thermodynamic equilibrium is achieved. The equilibrium hydrogen concentration in the grain boundary $C_\mathrm{H,GB,eq}$ can be calculated analytically using Equation~\eqref{eq:KKS}. Once equilibrium is reached, the amount of hydrogen trapped at the grain boundary $N_\mathrm{H,GB}$ is evaluated and should be equal to:
\begin{equation}
\begin{split}
N_\mathrm{H,GB} =& V_\mathrm{GB}C_\mathrm{H,GB,eq}\\
=& V_\mathrm{GB} \frac{f_\mathrm{lattice}}{V_\mathrm{m}}\frac{1}{1+\frac{1 - y_\mathrm{H,B} }{y_\mathrm{H,B}}\exp\left(\frac{\Delta \mu_\mathrm{H,b,GB}}{RT}\right)}. \\
\end{split}
\end{equation}
This procedure is then repeated for different grid spacings while keeping the same number of grid points across the diffuse interface (i.e., changing the resolution of the phase-field simulation). In doing so, the normalized interfacial thickness $\tilde{\lambda}$ (see Equation~\eqref{eq:normalized_interfacial_thickness}) is varied over several orders of magnitude, from $1$ up to $10^6$, effectively scaling the numerical phase-field interface thickness.

For a quantitatively correct model, the amount of hydrogen trapped at the GB must remain constant and be determined solely by the physical GB thickness and the trapping energy, independent of the numerical scaling. The results of this benchmark conducted in 1D, 2D, and 3D are shown in Figure~\ref{FIG:BM1}.
It is clear from Figure~\ref{FIG:BM1} that the GB hydrogen concentration remains invariant with respect to the interface scaling (from the true GB thickness ($\tilde{\lambda} = 1$), up to an interface thickness of six orders of magnitude larger ($\tilde{\lambda} = 10^6$)) and is independent of the dimensionality of the simulation (all curves overlap). 

\subsection{Benchmark 2: Equilibrium flux perpendicular to the grain boundary}
\begin{figure}[pos=h]
	\centering
		\includegraphics[width=\linewidth]{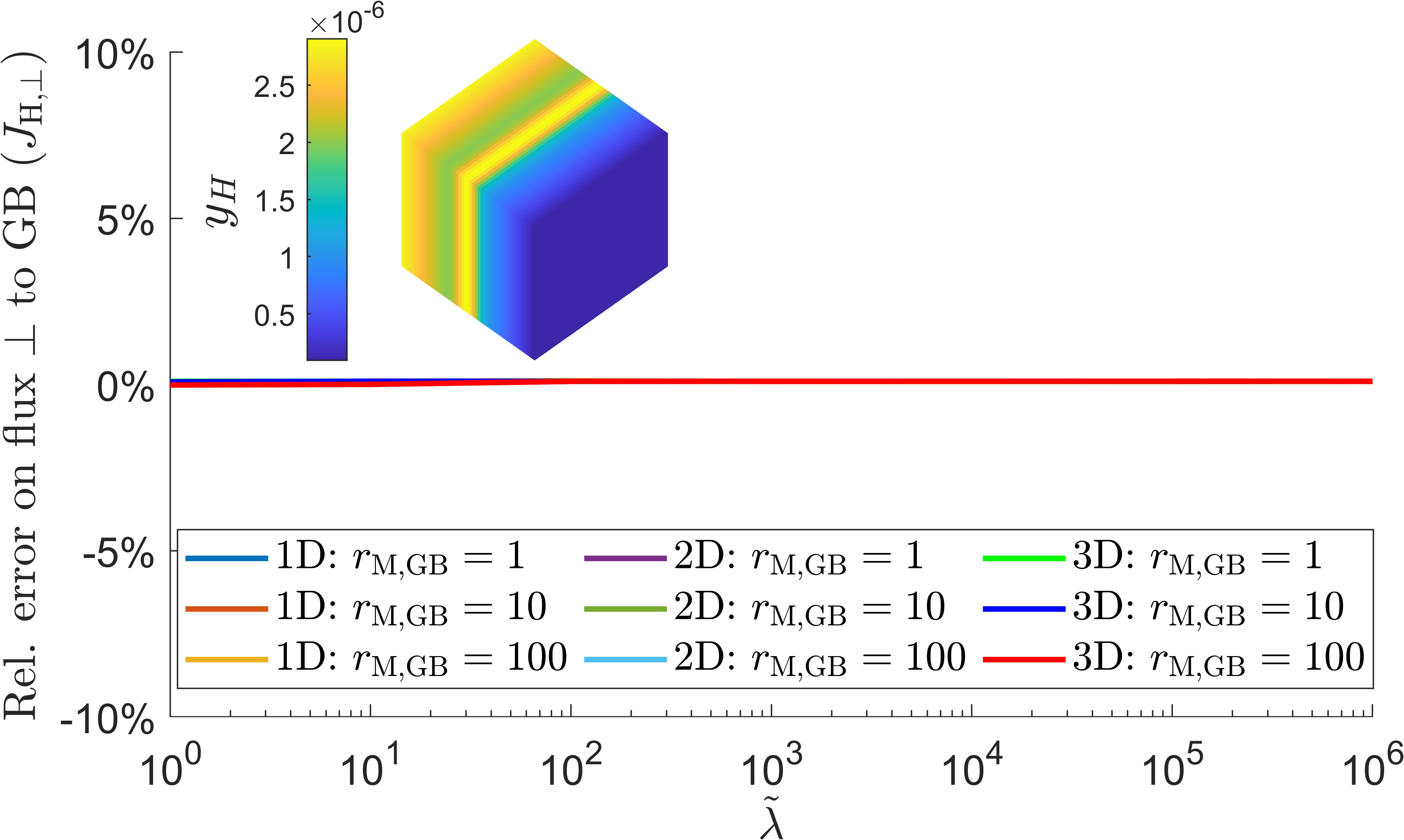}
	\caption{Benchmark 2: Relative error in the steady-state flux perpendicular to the GB compared to the analytical solution as a function of the normalized interfacial thickness.  The error remains below 1\% for all cases, indicating that all formulations are equivalent for practical applications.}
	\label{FIG:BM2}
\end{figure}

The second benchmark assesses whether the hydrogen flux perpendicular to a GB remains independent of the numerical interface scaling. Because physical GBs are only a few nanometers thick, transport across the boundary is assumed to be governed entirely by bulk diffusion. In other words, the time required for hydrogen to traverse the GB itself is negligible compared to the time required to diffuse through the surrounding bulk material. This behavior is enforced in the model through Equation~\eqref{eq:M_perp}. To rigorously test the formulation, the benchmark evaluates not only the effect of interface scaling and simulation dimensionality, but also the influence of the GB mobility. Physically, the GB mobility should not affect the flux perpendicular to the boundary, as dictated by Equation~\eqref{eq:M_perp}.
In the benchmark setup, a single GB is positioned at the center of the simulation domain, and a constant bulk hydrogen concentration gradient ($ \nabla C_\mathrm{H,B}$) is imposed across the system. The thermodynamic description of the hydrogen chemical potential corresponds to that of a regular solution model. As a result, an analytical expression for the steady-state flux, corresponding to Fick's first law \cite{Crank1975}, can be obtained using the bulk atomic mobility and concentration: 
\begin{equation}
\label{eq:flux_GB_perp}
\begin{split}
J_\mathrm{H,\perp}&= J_\mathrm{H,B}\\
&=-  M_\mathrm{at,H,B} C_\mathrm{H,B} \nabla \mu_\mathrm{H,B} \\
&=-  D_\mathrm{H,B} \nabla C_\mathrm{H,B} \\
\end{split}
\end{equation}
The numerical results are shown in Figure~\ref{FIG:BM2}.
As illustrated in Figure~\ref{FIG:BM2}, the computed flux is independent of the interface scaling, the dimensionality of the simulation, and the GB mobility (all curves overlap). 

\subsection{Benchmark 3: Equilibrium flux parallel to the grain boundary}
\begin{figure}[pos=h]
	\centering
		\includegraphics[width=\linewidth]{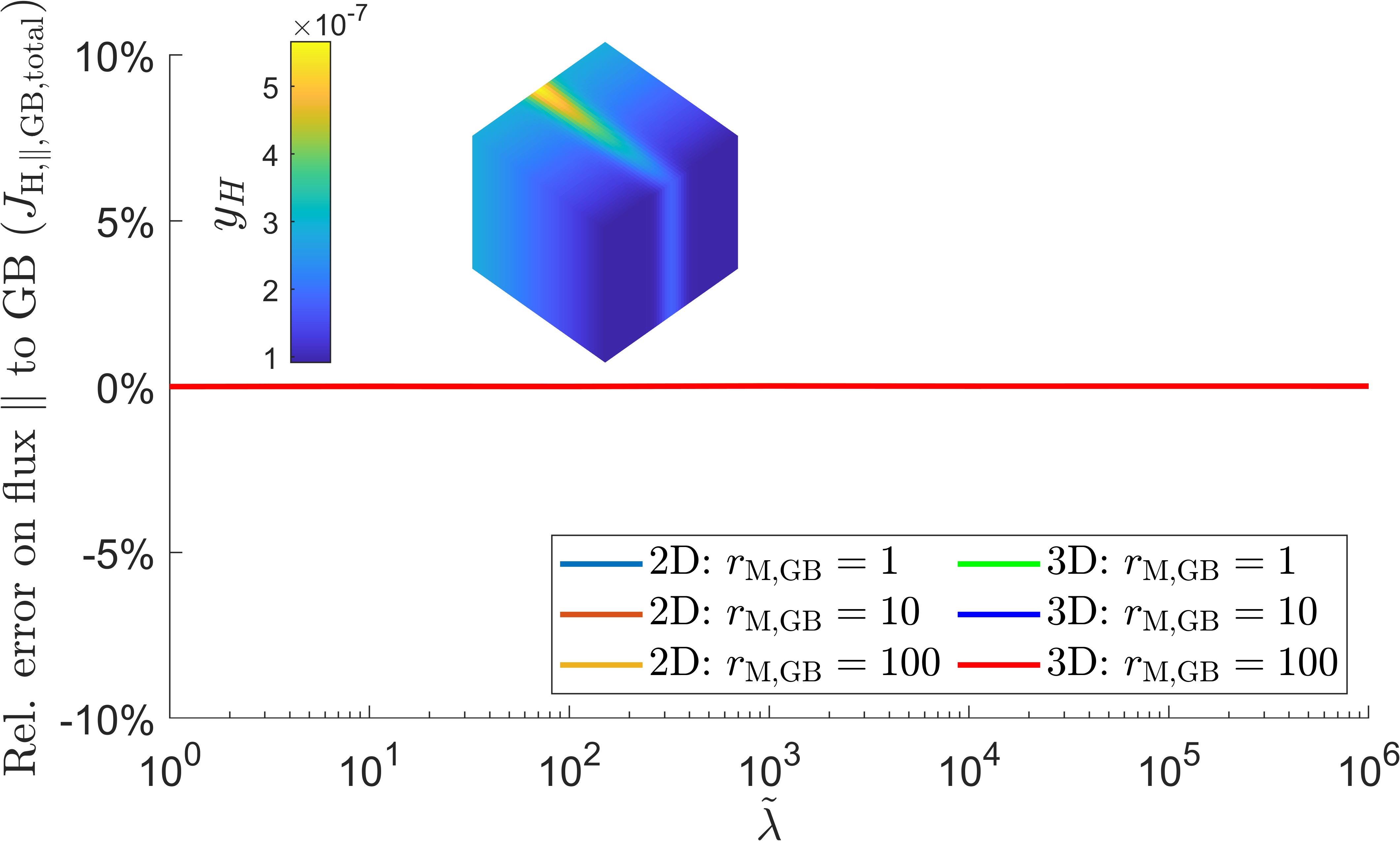}
	\caption{Benchmark 3: Relative error in the steady-state flux parallel to the GB compared to the analytical solution as a function of the normalized interfacial thickness. The error remains below 1\% for all cases, indicating that all formulations are equivalent for practical applications.}
	\label{FIG:BM3}
\end{figure}

The third benchmark naturally follows from Benchmark 2 and evaluates the steady-state hydrogen flux parallel to the GB. The total flux parallel to the GB ($J_\mathrm{H,\|,total}$) can be expressed as the sum of the bulk flux and GB flux. If the GB normal is aligned with the \(y\)-direction and the concentration gradient is imposed parallel to the GB plane, the total flux is:
\begin{equation}
\label{eq:flux_GB_para}
\begin{split}
J_\mathrm{H,\|,total} =& J_\mathrm{H,\|,B,total}+ J_\mathrm{H,\|,GB,total}\\
=& \iint J_\mathrm{H,B}\mathrm{d}x \mathrm{d}y + \iint J_\mathrm{H,GB}\mathrm{d}x\mathrm{d}y\\
=&-L_\mathrm{x}((L_\mathrm{y} - t_\mathrm{int,phys})M_\mathrm{at,H,B} C_\mathrm{H,B} \nabla \mu_\mathrm{H} \\
&+t_\mathrm{int,phys}M_\mathrm{at,H,GB} C_\mathrm{H,GB} \nabla \mu_\mathrm{H}) \\
=&-L_\mathrm{x}((L_\mathrm{y} - t_\mathrm{int,phys})D_\mathrm{H,B}\nabla C_\mathrm{H,B} \\
&+t_\mathrm{int,phys}D_\mathrm{H,GB} \nabla C_\mathrm{H,GB}). \\
\end{split}
\end{equation}
Here, the first term represents the total flux through the bulk of the material over the domain area (in the x-y-direction), while the second term accounts for diffusion along the GB over its physical thickness $t_\mathrm{int,phys}$. The relative error on the GB flux ($J_\mathrm{H,\|,GB,total}$) is shown in Figure~\ref{FIG:BM3}.
As demonstrated in Figure~\ref{FIG:BM3}, the numerically computed total flux parallel to the GB agrees with the analytical prediction for all tested interface thicknesses (varying $\tilde{\lambda}$), GB mobilities ($M_\mathrm{GB}$), and simulation dimensionalities (all curves overlap). 

Although the previous three benchmarks demonstrate that the steady-state hydrogen flux is captured correctly, it is also essential to validate the transient- or kinetic-behavior of the model. This is discussed in an additional benchmark in Appendix~C, omitted here for the sake of brevity.

\subsection{Benchmark 4: Upscaling desorption test}

A defining strength of a quantitative phase-field model is its ability to upscale the numerical interface thickness far beyond the nanometer-scale physical GB thickness while preserving the relevant nanoscale physics. This yields three key advantages. First, it enables the resolution of mesoscale microstructural features, such as individual grains in a polycrystal, while preserving the underlying nanoscale physics, such as GB trapping and diffusion. Second, increasing the numerical length scale, $t_\mathrm{int,PF}$, relaxes the stability constraint on the time step in the diffusion equation (Equation~\eqref{eq:pf_diffusion}), allowing for larger time steps and thus faster simulations. Third, a larger numerical length scale reduces the number of grid points required to represent a system of fixed physical size, further decreasing the computational cost. 

\begin{figure}[pos=h]
	\centering
		\includegraphics[width=\linewidth]{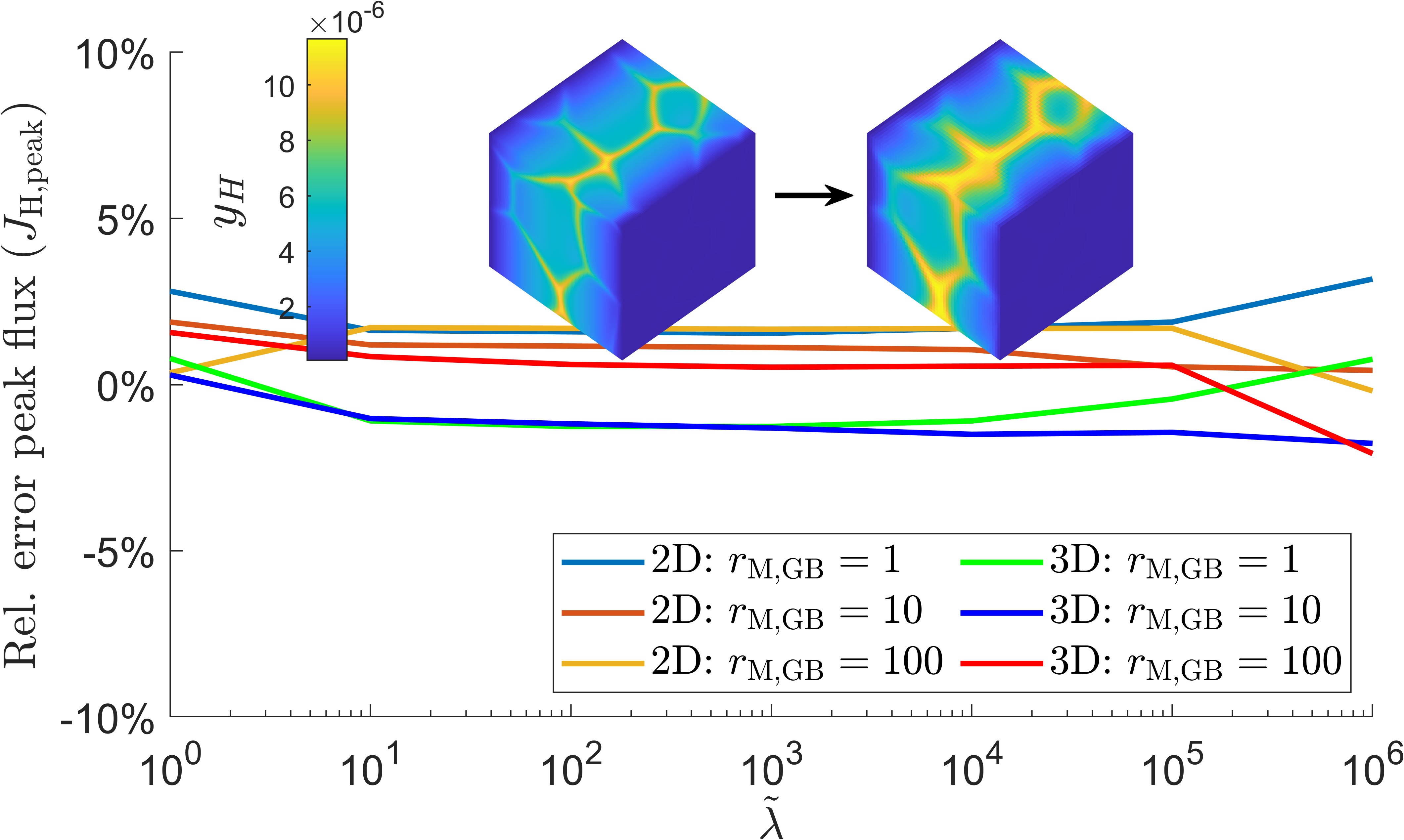}
	\caption{Benchmark 4: Relative error in the peak flux of the TDS spectrum, compared with the corresponding fine-mesh simulation, for different GB mobility ratios ($r_\mathrm{M,GB}=M_\mathrm{at,H,GB,0}/M_\mathrm{at,H,B,0}$) in two and three dimensions. The absolute errors remain below \(3\%\), reflecting the unavoidable loss of accuracy associated with the coarser spatial resolution. Nevertheless, as long as the grain structure is properly resolved, the formulation is identical for practical applications.}
	\label{FIG:BM5}
\end{figure}

In this benchmark, two simulations describing the same physical microstructure are compared. The first employs a grid of $100\times100\times100$ points containing nine grains (Figure~\ref{FIG:BM5}). The second represents the identical microstructure but uses a coarser grid of $50\times50\times50$ points, corresponding to a grid spacing that is twice as large, i.e., $t_\mathrm{int,PF,coarse}/t_\mathrm{int,PF,fine}=2$. This scaling factor cannot be chosen arbitrarily large, because both the diffuse interface and the microstructural features must remain numerically resolved. In particular, the diffuse-interface thickness should remain small compared with the smallest microstructural feature of interest \cite{NISTPhaseFieldRecommendedPractices}. Apart from the numerical resolution, both simulations describe the same physical system; consequently, their thermodynamic and kinetic trapping behavior should be identical.

To assess this, a thermal desorption spectroscopy (TDS) simulation is performed. The sample is first precharged with hydrogen and allowed to reach equilibrium. Vacuum boundary conditions are then applied at the sample surfaces in the $x$-direction, after which the temperature is increased at a constant heating rate until all hydrogen is released. This benchmark also probes the influence of triple junctions on hydrogen trapping, for which a compensation term was introduced in Equations~\ref{eq:h_B} and~\ref{eq:h_GB}. The primary quantity of interest is the peak of the TDS spectrum, corresponding to the maximum hydrogen flux ($J_\mathrm{H,peak}$) at a specific peak temperature ($T_\mathrm{peak}$), which is associated with detrapping from GBs. This peak is compared for the simulations with the coarse and the fine grid to estimate the relative error: $(J_\mathrm{H,peak,coarse} - J_\mathrm{H,peak,fine})/J_\mathrm{H,peak,fine}$.

The benchmark is repeated for different normalized interfacial widths of the fine meshed simulation ($\tilde{\lambda}_\mathrm{fine} = t_\mathrm{int,PF,fine}/t_\mathrm{int,phys}$) by varying the grid spacing, while the grid spacing of the coarse simulation is always kept twice as large. All simulations were conducted in 2D and 3D and for different GB mobility ratios. Figure~\ref{FIG:BM5} summarizes the results for all normalized interfacial thicknesses in both two- and three-dimensional simulations. The small absolute error, below \(3\%\), observed in the peak desorption flux confirms that nanoscale GB trapping and diffusion behavior are accurately preserved upon upscaling to physically relevant mesoscale resolutions. The residual error can be attributed to the unavoidable loss of accuracy associated with the coarser grid spacing: small grains and junctions are not reproduced perfectly, and the number of trapping sites therefore differs slightly between the fine and coarse simulations. At \(\tilde{\lambda}=1\), this difference is \(1.6\%\). This small error represents a favorable trade-off, since the coarse simulation reduces the computational time by a factor of 43.
\section{Application I: TDS interpretation}\label{sec:Model_results}
\subsection{Effect of grain boundary diffusion on TDS spectra}
\begin{figure}[pos=h]
	\centering
	\includegraphics[width=\linewidth]{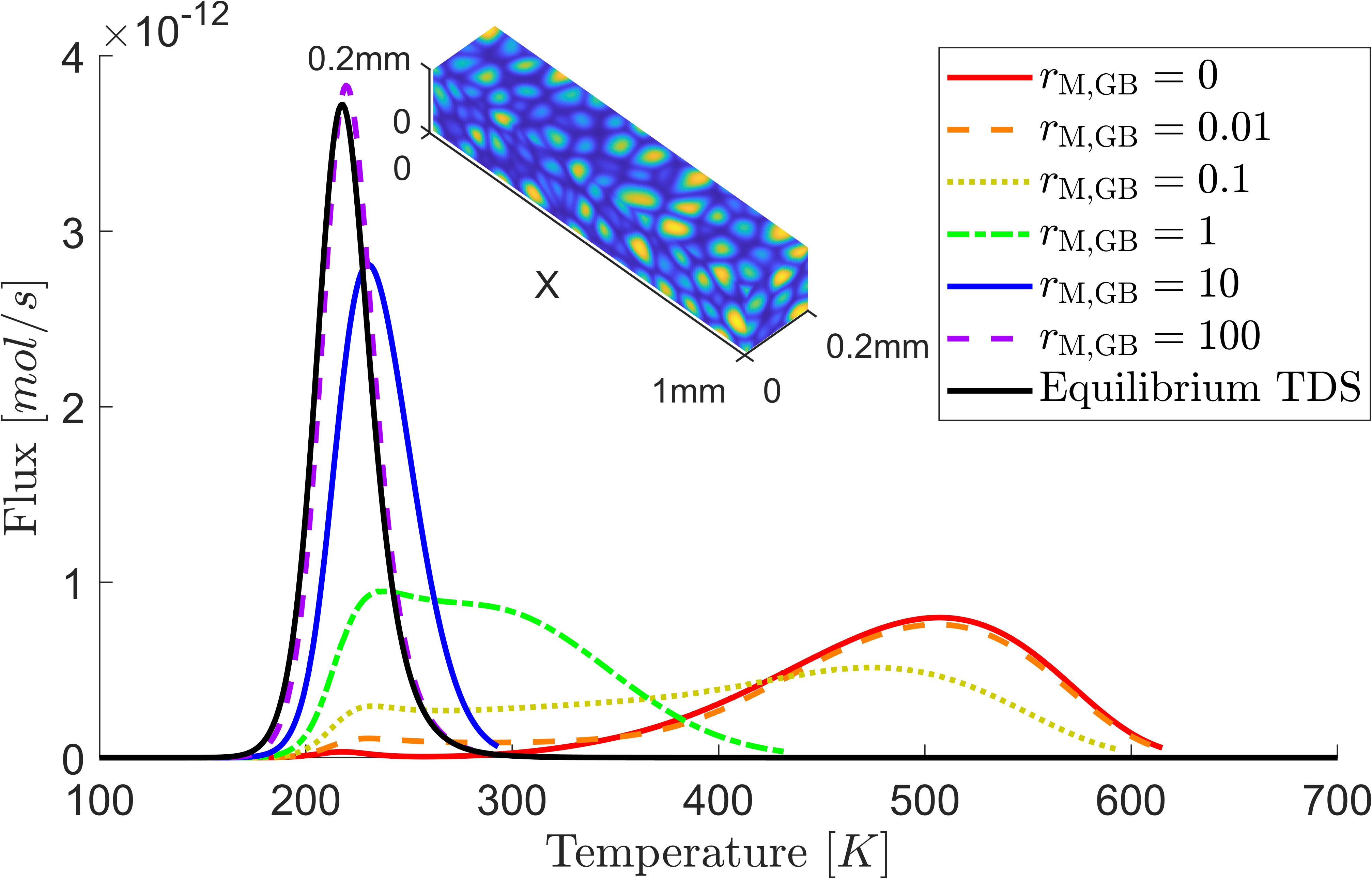}
    \caption{Cryogenic TDS spectra obtained for the same microstructure and different GB mobilities at a heating rate of \(\SI{5}{K.min^{-1}}\), illustrating the effect of varying GB mobility. The flux is shown only from \(100\,\mathrm{K}\) onward, since no measurable flux was observed below this temperature.}
	\label{FIG:Effect_of_M_GB_TDS_cryo}
\end{figure}

A key application of the proposed model, following validation of its quantitative behavior, is the investigation of how \textit{interconnected hydrogen traps}, GBs in particular, affect thermal desorption spectroscopy (TDS) spectra. To this end, three-dimensional simulations were performed on the same equiaxed microstructure (see the inset of Figure~\ref{FIG:Effect_of_M_GB_TDS_cryo}). In each case, the sample was first charged with hydrogen by means of gas charging. Subsequently, ultra-high vacuum (UHV) boundary conditions were applied at the surfaces normal to the \(x\)-direction, and the temperature was gradually increased while monitoring hydrogen desorption flux. The time between charging and TDS measurement under ultra-high vacuum (UHV) conditions was modeled to be negligible, avoiding H losses to the environment.

Since rapid diffusion along GBs may shift part of the desorption signal to temperatures below room temperature, cryogenic simulations were conducted, starting from \(\SI{50}{K}\). This allows the low-temperature desorption contribution associated with GB transport and trapping to be captured, in line with the motivation of cryogenic TDS experiments \cite{Sato2023}. Since cryo-TDS is experimentally more demanding and less commonly used, room temperature TDS simulations were also conducted with charging and desorption initiated at \(\SI{300}{K}\). All other parameters were kept identical. The baseline input parameters are summarized in Table~\ref{tab:TDS_parameters}; further details are provided in the input parameter scripts \cite{feyen2026_MendeleyData}. 
\begin{table}[pos=h]
\centering
\caption{Main input parameters used in the TDS simulations.}
\label{tab:TDS_parameters}
\begin{tabularx}{\columnwidth}{@{}p{0.56\columnwidth} X@{}}
\toprule
\textbf{Parameter} & \textbf{Value} \\
\midrule
Sublattice fraction $f_\mathrm{lattice}$ & \(6\) \\
% At. bulk mobility $M_\mathrm{at,H,B,0}$ & $\frac{1.379 \times 10^{-8}}{RT}\exp\left(\frac{-9312}{RT}\right)$ $\frac{m^2mol}{Js}$~\cite{He2017}\\
Bulk migration energy $E_\mathrm{migr}$ & \(\SI{9312}{J.mol^{-1}}\)~\cite{He2017}\\
Bulk diffusion prefactor $D_\mathrm{B,0}$ & \(\SI{1.379e-8}{m^2.s^{-1}}\)~\cite{He2017}\\
GB mobility ratio $r_\mathrm{M,GB}$ & \(0, 0.01, 0.1, 1, 10, 100\)\\
Molar volume \(V_m\) & \(\SI{7.09e-6}{m^3.mol^{-1}}\)~\cite{Singman1984} \\
GB binding free energy \(\Delta \mu_\mathrm{H,b,GB}\) & \(\SI{-55}{kJ.mol^{-1}}\) \\
Plate thickness, \(L_x\) & $1$ mm\\
Grid size $nx \cross ny \cross nz$ & \(200 \times 40 \times 40\) \\
Grid spacing \(\Delta x\) & $5$ \textmu m \\
Physical interface thickness \(t_\mathrm{int,phys}\) & $1$ nm~\cite{Hamza2015} \\
PF interfacial thickness \(t_\mathrm{int,PF}\) & $30$ \textmu m\\
Normalized interfacial thickness \(\tilde{\lambda}\) & \(\SI{3e4}{}\) \\
Number of grains $n_\mathrm{grains}$ & \(117\) \\
Average grain size $d_\mathrm{grain}$& $87$ \textmu m \\
H Charging pressure $P_\mathrm{Char}$& \(\SI{1e7}{Pa}\) \\
Charging temp. $T_\mathrm{Char}$ & \(50, \SI{300}{K}\) \\
Desorption pressure $P_\mathrm{UHV}$ & \(\SI{2.5e-7}{Pa}\) \\
Heating rates $\dot{T}$ & \(1,5,10, \SI{15}{K.min^{-1}}\) \\
Dirichlet boundary conditions & \(x\)-direction \\
Periodic boundary conditions & \(y\)-, \(z\)-direction \\
\bottomrule
\end{tabularx}
\end{table}
The parameter set is representative of BCC iron, highly relevant for the pipeline and pressure-vessel applications, but the aim is neither to reproduce the response of a specific material system nor to mimic a realistic experimental TDS protocol \cite{Fangnon2020}. Instead, the simulations are designed to isolate the general effect of finite GB mobility on idealized TDS spectra. The literature reports a broad range of values for the GB binding free energy, reflecting the strong dependence of hydrogen trapping on the atomic structure of the grain boundary \cite{Starikov2022, Hamza2015, MATSUMOTO2011, Sato2023}. For illustrative purposes, the GB binding free energy (\(\Delta \mu_\mathrm{H,b,GB} = -E_\mathrm{trap}\)) was assumed to be temperature independent and was set to \(\SI{-55}{kJ.mol^{-1}}\). This value lies within the reported range, and additional parametric studies showed that the main conclusions drawn here remain qualitatively similar for other trapping energies.

The resulting cryogenic TDS spectra for a heating rate of \(\SI{5}{K.min^{-1}}\) are shown in Figure~\ref{FIG:Effect_of_M_GB_TDS_cryo}. In addition to the simulated TDS spectra, an equilibrium TDS curve is included. This curve represents the limiting response that would be obtained if the sample remained in equilibrium with the UHV gas phase at each temperature during the TDS experiment. In this limit, Oriani's local equilibrium assumption effectively becomes a global equilibrium condition, or equivalently, detrapping and transport kinetics are no longer rate-limiting. The derivation of the equilibrium TDS curve is given in Appendix~D.

When the GB mobility is zero (i.e., $r_\mathrm{M,GB}=0$), the spectrum contains two peaks: a small low-temperature peak, corresponding to the initial diffusion of lattice-dissolved hydrogen out of the sample, and a larger peak, corresponding to hydrogen release from GB traps. This situation corresponds to the grain boundary acting as an isolated trap meaning that no transport parallel to the GB can occur. As  $r_\mathrm{M,GB}$ is increased, both peaks begin to shift. Up to $r_\mathrm{M,GB}=0.01$, the GB peak remains nearly identical to the zero-mobility case, while the lattice peak increases slightly. The latter effect is caused by the GB network facilitating hydrogen transport out of the sample. When  $r_\mathrm{M,GB}$ reaches \(0.1\), the GB peak decreases in intensity and its local maximum shifts to lower temperatures. When the GB mobility equals the bulk mobility, the two peaks overlap significantly. For  $r_\mathrm{M,GB}=10$ the two peaks merge into a single peak. In this regime, hydrogen release from GB traps occurs simultaneously with the initial removal of lattice-dissolved hydrogen. The apparent desorption peak, therefore, becomes controlled by rapid transport along the interconnected GB network rather than by bulk diffusion away from isolated traps. This effect becomes even more pronounced for \(r_\mathrm{M,GB}=100\). In this case, the spectrum is nearly identical to the equilibrium TDS curve, indicating that detrapping and transport kinetics are no longer rate-limiting at the heating rate used in the simulation.

\begin{figure}[pos=h]
	\centering
	\includegraphics[width=\linewidth]{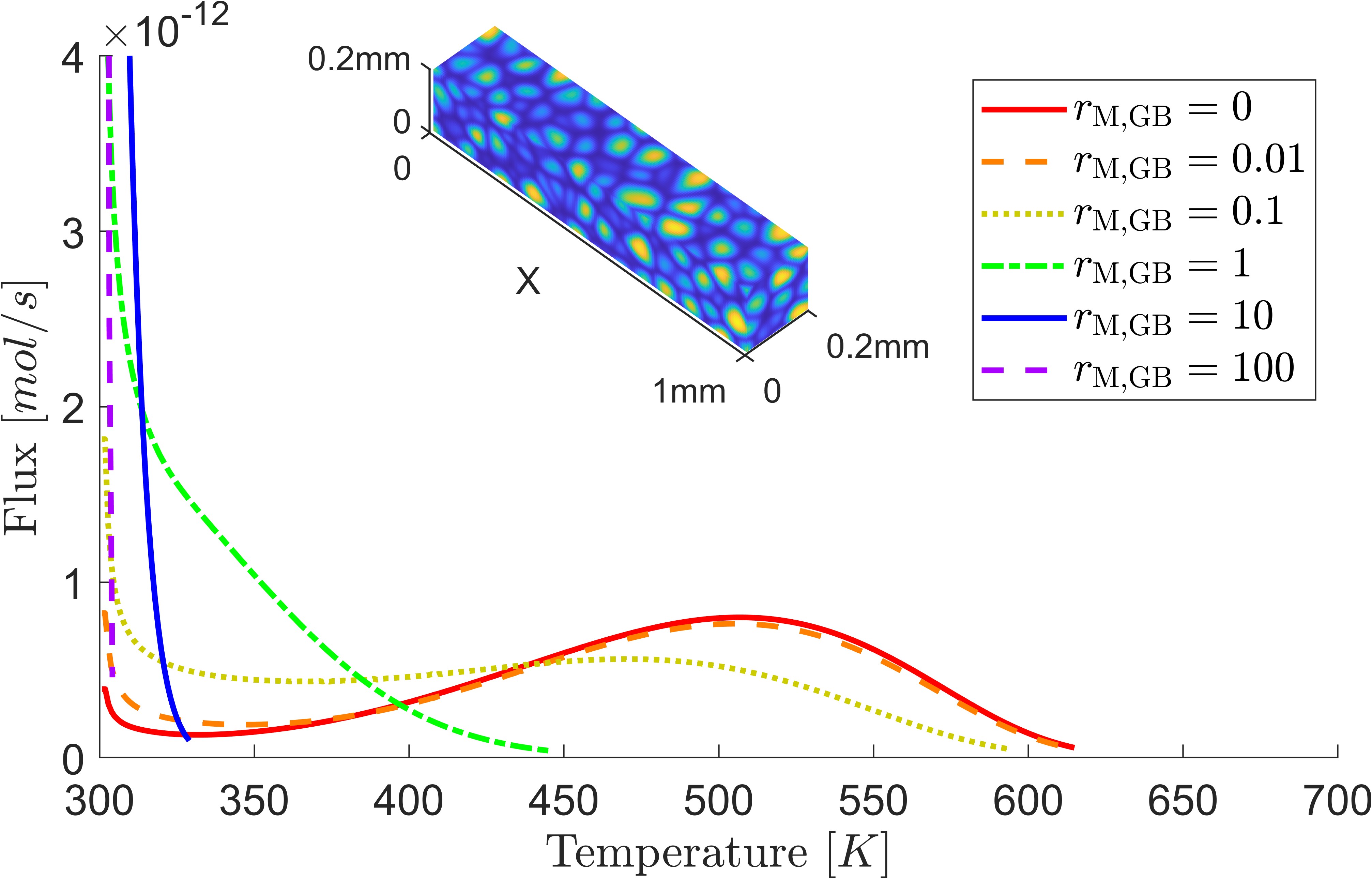}
    \caption{Room-temperature TDS spectra obtained for the same microstructure and different GB mobilities at a heating rate of \(\SI{5}{K.min^{-1}}\). The spectra illustrate the combined effect of the starting temperature and GB mobility on the TDS response. At high GB mobility, the GB detrapping peak is no longer visible.}
	\label{FIG:Effect_of_M_GB_TDS}
\end{figure}

The resulting room-temperature TDS spectra are shown in Figure~\ref{FIG:Effect_of_M_GB_TDS}. Although a trapping energy magnitude of \(\SI{55}{kJ.mol^{-1}}\) is usually considered characteristic of a relatively deep trap, sufficiently high GB mobility can shift the main desorption contribution below room temperature, making the corresponding peak no longer detectable in conventional TDS. A similar effect may occur for other traps that are sufficiently coupled to the GB network, since fast GB transport can accelerate the removal of hydrogen released from nearby trapping sites.

Even when \(M_\mathrm{at,H,GB}=M_\mathrm{at,H,B}\), the effective parallel transport along the GB is enhanced because the flux contribution scales with \(M_\mathrm{at,H,GB}y_\mathrm{H,GB}\), as shown in Equation~\ref{eq:M_para}. Since \(y_\mathrm{H,GB}\gg y_\mathrm{H,B}\) for deep hydrogen traps, the GB can contribute significantly to the total flux even when the intrinsic atomic mobility is not larger than in the bulk.

These findings have important implications for the interpretation of experimental TDS spectra. In particular, they demonstrate that GB diffusion can significantly alter peak positions and peak shapes, potentially leading to misinterpretation of trapping energies if diffusion effects are not properly accounted for. Consequently, TDS peak positions cannot be interpreted solely in terms of trap binding energy when traps form an interconnected transport network. In such cases, the measured peak temperature reflects the coupled influence of trapping thermodynamics, bulk diffusion, and GB-assisted transport. This issue is examined in more detail in the following section. 

\subsection{Kissinger analysis of TDS spectra with grain boundary diffusion}
Another important application of the developed modeling framework is its use for inferring GB trapping energies from TDS spectra. Traditionally, the Kissinger equation is widely used to estimate trapping energies from TDS experiments. This approach relies on repeated hydrogen charging and desorption experiments conducted at different heating rates, $\dot{T}$. The resulting peak temperatures, $T_\mathrm{peak}$, are then analyzed using a Choo-Lee or Kissinger-type plot to extract an apparent activation energy \cite{Blaine2012, Choo1982}:
\begin{equation}
\label{eq:Kissinger}
\begin{split}
\frac{\partial \ln\left(\dot{T}/T_\mathrm{peak}^2\right)}
{\partial\left(1/T_\mathrm{peak}\right)}
&= -\frac{E_\mathrm{a}}{R} \\
&= -\frac{E_\mathrm{trap}+E_\mathrm{migr}}{R}.
\end{split}
\end{equation}
Here, \(E_\mathrm{a}\) denotes the apparent activation energy associated with the desorption peak. In the classical interpretation, this activation energy is related to the sum of the trap binding energy, \(E_\mathrm{trap}\), and the migration energy for lattice diffusion, \(E_\mathrm{migr}\) (=9312 J/mol in this simulation set, see Table \ref{tab:TDS_parameters}).

By performing simulations with a known trapping energy magnitude, \(E_\mathrm{trap,true}=\SI{55}{kJ.mol^{-1}}\), and varying the heating rate as summarized in Table~\ref{tab:TDS_parameters}, the Kissinger analysis can be applied directly to the simulated spectra. This makes it possible to assess whether the trapping energy inferred from the Kissinger equation (\(E_\mathrm{trap,est}\)) corresponds to the actual trapping energy imposed in the simulations (\(E_\mathrm{trap,true}\)).

\begin{figure}[pos=h]
	\centering
		\includegraphics[width=\linewidth]{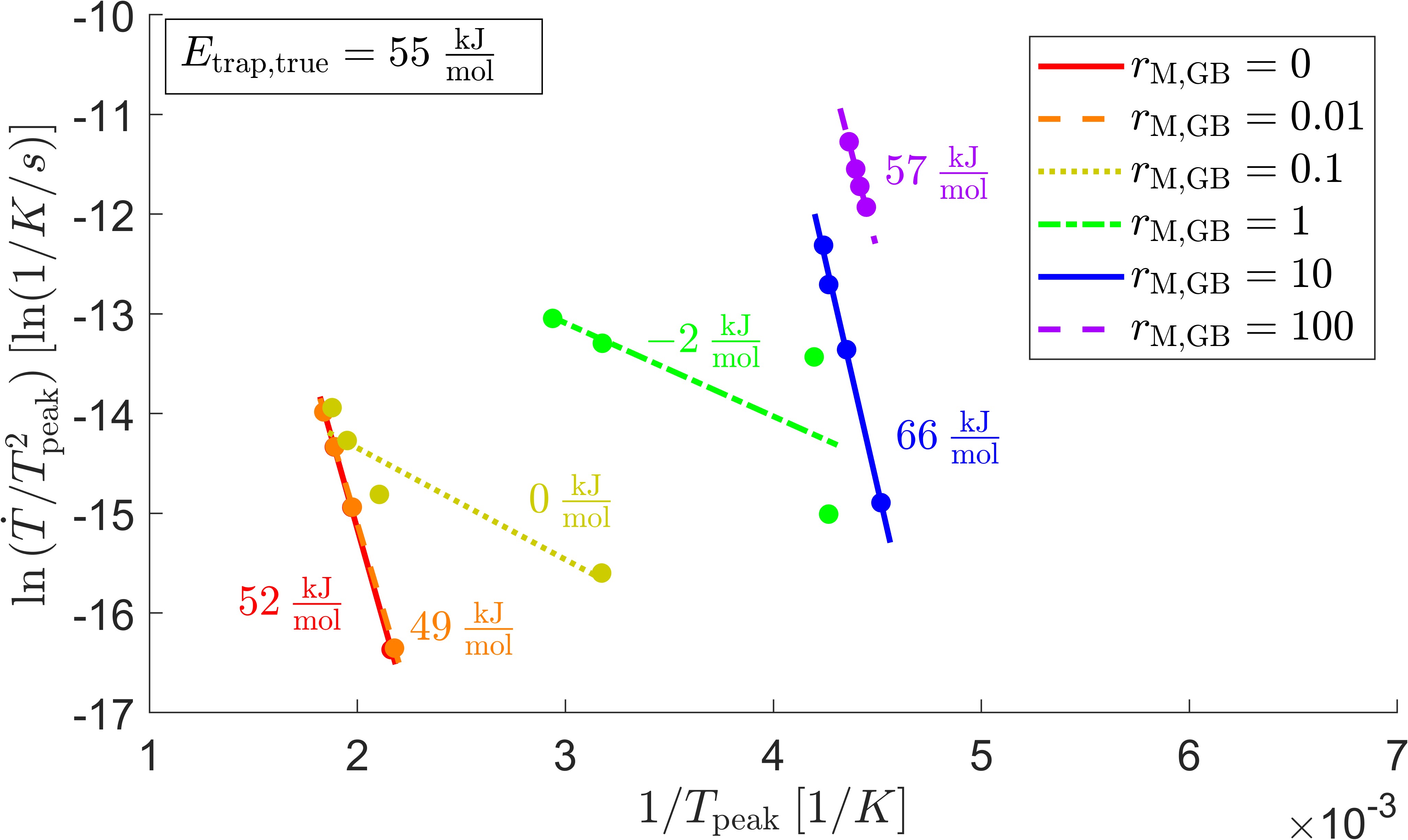}
	\caption{Choo-Lee plot for simulations with identical GB trapping energy but different GB mobilities. The trapping energies estimated from the fitted slopes (\(E_\mathrm{trap,est}\)) are indicated for each curve. Only at very low or very high grain boundary mobility, the correct trapping energy is recovered.}
	\label{FIG:Effect_of_M_GB_Kiss}
\end{figure}

The Choo-Lee plot~\cite{Choo1982}, showing \(\ln(\dot{T}/T_\mathrm{peak}^2)\) as a function of \(1/T_\mathrm{peak}\), is depicted in Figure~\ref{FIG:Effect_of_M_GB_Kiss}. The rates for \(r_\mathrm{M,GB} =  100\), were selected to be equal to $20,25,30,\SI{40}{K.min^{-1}}$. They deviate from those listed in Table \ref{tab:TDS_parameters}, since for these slower heating rates the spectra were too close to equilibrium, making it impossible to derive activation energies from them. 

The trapping energy can be estimated from the slope of the fitted curves shown in Figure~\ref{FIG:Effect_of_M_GB_Kiss}. For \(r_\mathrm{M,GB}\leq 0.01\), detrapping is dominated by bulk diffusion, and the trapping energy estimated using the Kissinger equation agrees with the imposed value of \(\SI{55}{kJ.mol^{-1}}\) within numerical accuracy. At higher ratios, however, GB-assisted transport contributes significantly to the desorption process. In this regime, the assumptions underlying the classical Kissinger interpretation are no longer satisfied, and the extracted trapping energy deviates strongly from the imposed value. For \(0.1\leq r_\mathrm{M,GB}\leq 1\), a clear linear relation can no longer be obtained from the peak temperatures because the corresponding peaks begin to overlap, depending on the heating rate, as illustrated in Figure~\ref{FIG:Effect_of_M_GB_TDS_cryo} for \(r_\mathrm{M,GB}=1\).

It is important to note that only \(r_\mathrm{M,GB}\) was varied in this set of simulations, while the temperature dependence of the GB mobility was kept unchanged. Therefore, changes in the Kissinger slope do not originate from a different prescribed activation energy for diffusion, but from changes in the mechanism controlling the peak position, including overlap between the lattice- and GB-related desorption contributions. In the limiting cases where either bulk diffusion (\(r_\mathrm{M,GB} = 0\)) or GB-assisted diffusion dominates (\(r_\mathrm{M,GB} = 100\)) over the full heating-rate range, the fitted slopes remain internally consistent. In the intermediate regime, however, the peak position reflects a changing combination of detrapping, bulk diffusion, and GB-assisted transport, leading to erroneous activation-energy estimates.

These results indicate that applying the Kissinger equation to experimental TDS data without accounting for GB diffusion can lead to substantial errors in the inferred trapping energies. The present modeling framework therefore provides a more reliable alternative for interpreting TDS experiments when GB diffusion plays a significant role. Instead of relying solely on peak positions, the framework enables trapping and transport parameters to be inferred by directly simulating the TDS experiment and comparing the resulting spectra with experimental measurements.

It is worth noting that molecular dynamics and DFT studies have shown that the relative magnitude of hydrogen mobility along GBs depends strongly on the crystal structure \cite{Smirnova2023, Hamza2015, Zhou2019}. For some BCC iron GBs, hydrogen diffusion along the boundary has been reported to be slower than bulk diffusion~\cite{Smirnova2023, Hamza2015, Zhou2019}, whereas in some FCC systems, such as Ni-based systems~\cite{Zhou2019}, or Fe-based systems~\cite{Smirnova2023}, GBs may provide faster diffusion pathways. This reinforces the need for material- and microstructure-specific interpretation of TDS spectra.

\section{Application II: Effective diffusion in polycrystals}

Analytical treatments, numerical models and experimental studies of diffusion with trapping commonly introduce an effective diffusion coefficient to describe the apparent rate of hydrogen transport in alloys \cite{Oriani1970, McNabb1963, ASTM_G0148_97R18, Zafra2022, Drexler2021, LopesPinto2024}. In this section, we examine how the presence of interconnected trapping sites, in particular GBs, modifies the effective diffusion behavior at the macroscopic scale. To this end, a systematic parametric study was performed in which the average grain size ($d_\mathrm{grain}$), GB binding free energy (\(\Delta \mu_\mathrm{H,b,GB}\)), temperature, and GB mobility of H were independently varied.

\begin{figure}[pos=h]
	\centering
		\includegraphics[width=\linewidth]{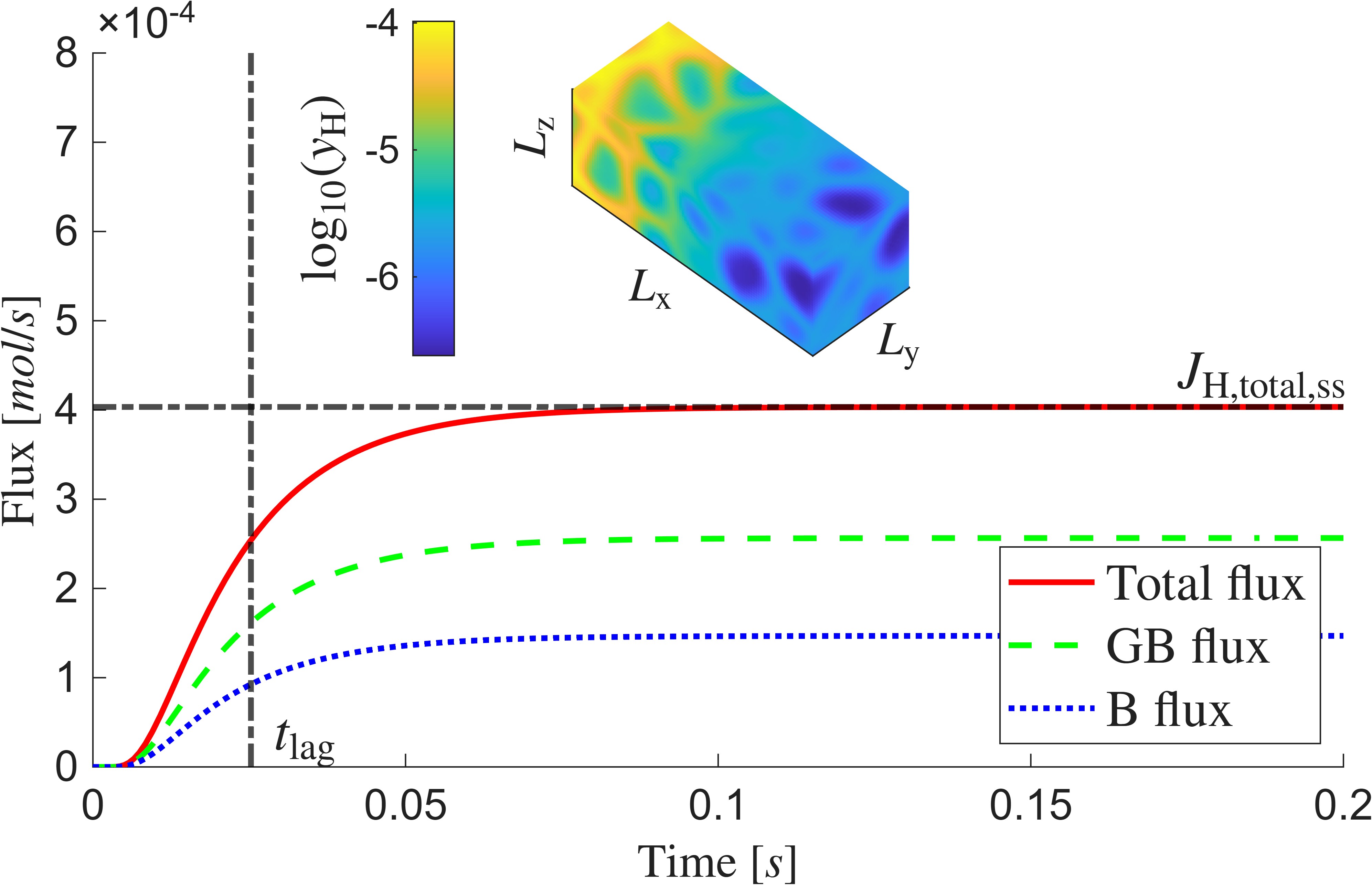}
        \caption{Illustration of the simulated hydrogen permeation response. The total hydrogen flux is decomposed into the GB and bulk contributions. The time lag, \(t_{\mathrm{lag}}\), is determined at 63\% of the steady-state flux level \(J_{\mathrm{H,total,ss}}\). The inset shows the corresponding three-dimensional hydrogen concentration field, expressed as \(\log_{10}(y_\mathrm{H})\), in the simulated polycrystalline microstructure.}
	\label{FIG:Permeation}
\end{figure}

All simulations were conducted on a representative three-dimensional microstructure with equiaxed grains, discretized on a $100 \times 40 \times 40$ grid containing 60 grains, illustrated in Figure \ref{FIG:Permeation}. $\Delta x$ was varied to obtain different grain sizes. A small gradient in bulk hydrogen concentration was imposed along the $x$‑direction (Dirichlet boundary conditions), while zero‑flux (Neumann) boundary conditions were applied in the $y$‑ and $z$‑directions.

According to the ASTM~G148 standard \cite{ASTM_G0148_97R18}, the hydrogen flux measured during an electrochemical permeation experiment can be related to an effective diffusion coefficient using two complementary approaches. The first relies on the steady‑state permeation flux, whereas the second extracts the diffusion coefficient from the transient permeation response \cite{ASTM_G0148_97R18}.

We first define the \textit{effective steady-state diffusion coefficient} $D_\mathrm{eff,ss}$, obtained from the steady-state hydrogen flux $J_\mathrm{H,total,ss}$ as
\begin{equation}
\label{eq:D_eff_ss}
D_\mathrm{eff,ss} = -\frac{J_\mathrm{H,total,ss} L_x}{L_\mathrm{y}L_\mathrm{z}(C_\mathrm{H,B,x=0} - C_\mathrm{H,B,x=L_x})},
\end{equation}
where $L_x,L_y,L_z$ is the length of the sample in the x, y, and z-direction, respectively (see Figure \ref{FIG:Permeation}). $C_\mathrm{H,B,x=0}$ and $C_\mathrm{H,B,x=L_x}$ denote the subsurface bulk hydrogen concentrations at the two ends of the sample.

In the transient permeation method, the effective diffusion coefficient is determined from the time evolution of the outgoing hydrogen flux. Specifically, the lag time $t_\mathrm{lag}$ is defined as the time required for the flux to reach 63\% of its steady-state value, i.e.\ $J(t)/J_\mathrm{H, total, ss} = 0.63$~\cite{ASTM_G0148_97R18} (see Figure \ref{FIG:Permeation}). The corresponding \textit{effective transient-state diffusion coefficient} $D_\mathrm{eff,ts}$ is then given by
\begin{equation}
\label{eq:D_eff_ts}
D_\mathrm{eff,ts} = \frac{L_x^2}{6 t_\mathrm{lag}}.
\end{equation}

In addition, an analytical expression for the effective diffusion coefficient in the presence of hydrogen traps was originally proposed by Oriani \cite{Oriani1970}. Oriani’s formulation assumes non‑interconnected traps and therefore accounts exclusively for bulk diffusion. In the notation adopted here, and only considering the grain boundaries as trapping sites \textit{Oriani’s effective diffusion coefficient} $D_\mathrm{eff,Or}$ can be written as (see Appendix~E)
\begin{equation}
\label{eq:D_eff_or}
D_\mathrm{eff,Or} \approx D_\mathrm{H,B} 
\frac{1}{1 + \dfrac{y_\mathrm{H,GB} V_\mathrm{GB}}{y_\mathrm{H,B} (V - V_\mathrm{GB})} (1 - y_\mathrm{H,GB})},
\end{equation}
where $V$ and $V_\mathrm{GB}$ are the total system volume and the total GB volume, respectively.

\begin{figure}[pos=h]
	\centering
		\includegraphics[width=\linewidth]{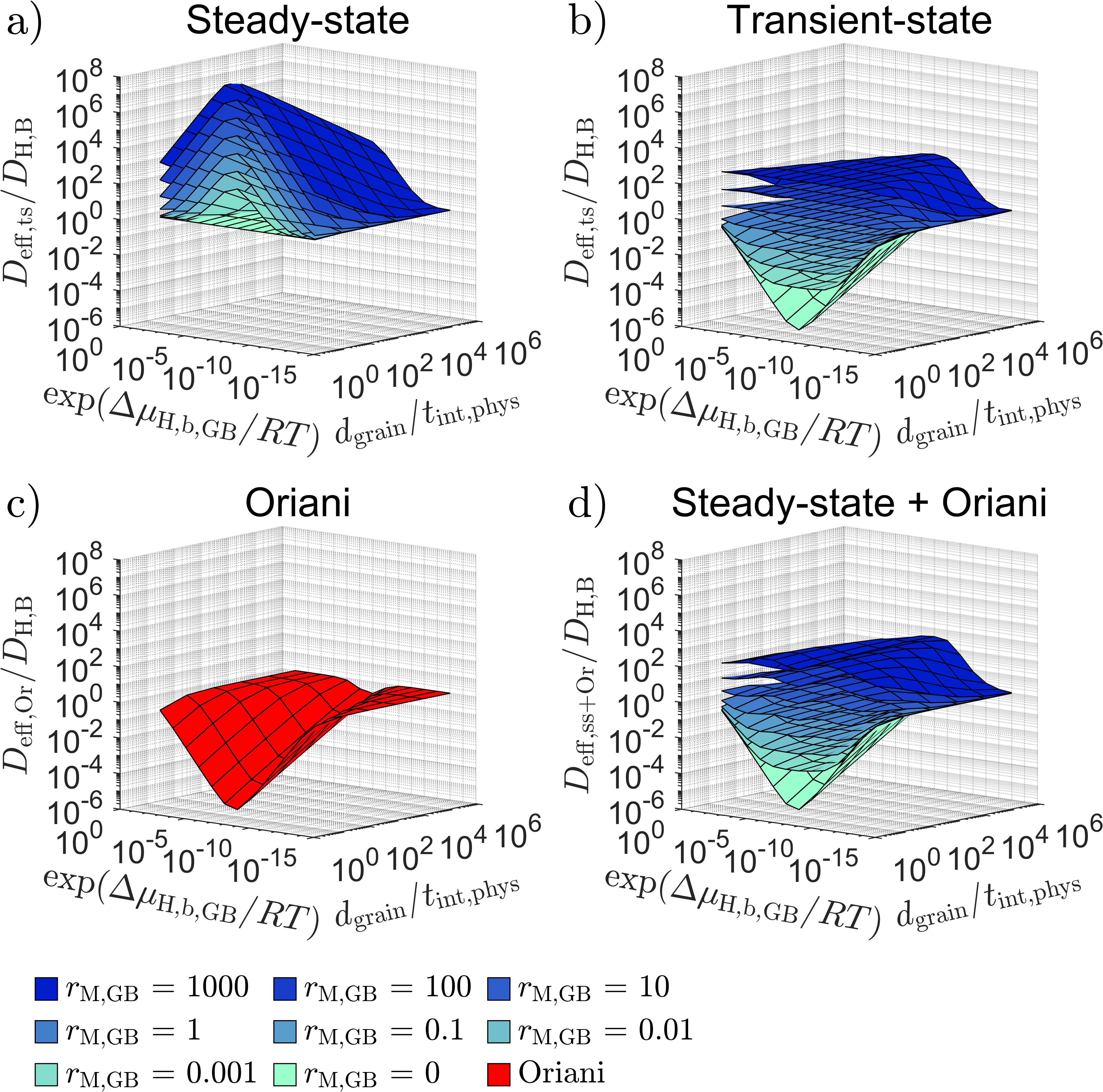}
	\caption{Non-dimensionalized steady-state (a), transient-state (b), and Oriani (c) effective diffusion coefficients, for a varying GB mobility ratio, average grain sizes, temperatures, and trapping energies. (d) Depicts the logarithmic sum of the steady-state and Oriani effective diffusion coefficient. The 3D plots are intended mainly as illustrative overviews; detailed contour plots for each value of \(r_\mathrm{M,GB}\) are provided in the supplementary materials.}
	\label{FIG:D_eff}
\end{figure}

For each simulation, all three effective diffusion coefficients were evaluated. The results are presented in non‑dimensional form to ensure generality across material systems with equiaxed, single-phase microstructures. Figures~\ref{FIG:D_eff}a-c show the steady-state, transient-state, and Oriani effective diffusion coefficient, respectively.

The steady-state diffusion coefficient $D_\mathrm{eff, ss}$ (Figure~\ref{FIG:D_eff}a) increases with decreasing grain size and increasing GB mobility, reflecting the growing contribution of fast GB transport pathways. A non‑monotonic dependence on trapping free energy and temperature is also observed. At high trapping energies, GB sites become nearly saturated, reducing the availability of interstitial vacancies and thus diminishing GB diffusivity. Conversely, at very low trapping energies, the GB hydrogen concentration remains small, leading again to a limited contribution to the total flux. As a consequence, $D_\mathrm{eff, ss}$ exhibits a maximum at intermediate trapping energies and temperatures. When the GB mobility is set to zero, $D_\mathrm{eff, ss}$ correctly reduces to the bulk diffusion coefficient $D_\mathrm{H,B}$. Notably, even for $r_\mathrm{M,GB} = 10^{-3}$, GBs already provide a substantial contribution to the steady-state flux, at intermediate trapping energies and temperatures.

The effective transient-state diffusion coefficient $D_\mathrm{eff,ts}$ (Figure~\ref{FIG:D_eff}b) is consistently lower than $D_\mathrm{eff,ss}$. This reduction arises from the finite time required to fill the GB traps. This effect is captured by Oriani’s effective diffusion coefficient (Figure~\ref{FIG:D_eff}c), which is always smaller than $D_\mathrm{H,B}$ and exhibits its strongest reduction at small grain sizes and intermediate trapping energies and temperatures. In this regime, traps are sufficiently strong to delay diffusion, yet not so strong that they become rapidly saturated.

The relationship between these three effective diffusion coefficients is elucidated in Figure~\ref{FIG:D_eff}d, which shows the logarithmic sum of $D_\mathrm{eff,ss}$ and $D_\mathrm{eff,Or}$. Comparison with Figure~\ref{FIG:D_eff}b reveals the approximate relation:
\begin{equation}
\label{eq:D_eff_relation}
\log_{10}\left(\frac{D_\mathrm{eff,ts}}{D_\mathrm{H,B}}\right)
\approx
\log_{10}\left(\frac{D_\mathrm{eff,ss}}{D_\mathrm{H,B}}\right)
+
\log_{10}\left(\frac{D_\mathrm{eff,Or}}{D_\mathrm{H,B}}\right).
\end{equation}
This scaling indicates that the delay associated with trap filling is partially compensated by enhanced long-range transport through the GB network. In this interpretation, $D_\mathrm{eff,Or}$ quantifies the retarding effect of trapping, while $D_\mathrm{eff,ss}$ captures the contribution of GB-assisted diffusion. Their combined effect governs the transient permeation response. Consequently, discrepancies between $D_\mathrm{eff,ss}$ and $D_\mathrm{eff,ts}$ provide direct insight into trapping kinetics, whereas the absolute magnitude of $D_\mathrm{eff,ss}$ reflects the efficiency of GB transport. A remaining experimental challenge lies in the accurate determination of the subsurface concentrations $C_\mathrm{H,B,x=0}$ and $C_\mathrm{H,B,x=L_x}$.

Finally, Appendix~F presents a fitted analytical expression for $D_\mathrm{eff,ss}$ based on six non-dimensional coefficients. Owing to its fully non‑dimensional form, this relation can be readily applied to a wide range of materials with equiaxed, single-phase microstructures. Experimental validation of this relation will be addressed in future work.

\section{Conclusions}
In this work, a quantitative phase-field model was developed to describe hydrogen diffusion, GB trapping, and GB-assisted transport in polycrystalline alloys. By combining thermodynamic and kinetic input parameters with an explicit phase-field representation of the microstructure, the model enables the coupled effects of trapping, bulk diffusion, and GB-assisted transport to be quantified. The main conclusions are as follows:

\begin{itemize}
    \item GBs affect hydrogen transport not only by acting as trapping sites, but also by forming interconnected diffusion pathways. When the GB mobility is non-negligible relative to the bulk mobility, even if it is several orders of magnitude smaller, hydrogen transport and release kinetics are governed by a coupled interplay between detrapping, bulk diffusion, and GB-assisted diffusion. Classical diffusion-trapping models that assume isolated traps may therefore be insufficient to describe the diffusion behavior. In such cases, the connectivity of the trap network must be considered explicitly, as enabled by the proposed framework.

    \item The proposed phase-field formulation is quantitative with respect to numerical interface scaling. It preserves the physical GB volume and the associated number of trapping sites independently of the chosen diffuse-interface thickness. Benchmark simulations confirmed that equilibrium trapping, fluxes perpendicular and parallel to GBs, and upscaled desorption behavior remain invariant with respect to numerical interface scaling. This distinguishes the formulation from existing phase-field models in which the numerical interface thickness influences the predicted transport behavior.
    
    \item A novel set of benchmark problems is proposed for phase-field models that describe grain boundary diffusion and trapping.

    \item TDS simulations showed that GB diffusion can significantly shift peak positions and alter peak shapes. As a result, Kissinger-type analyses yield reliable trapping energies only when GB diffusion does not contribute appreciably to hydrogen transport. Even modest GB mobility can lead to systematic errors in the inferred trapping energy.

    \item GB-assisted transport can shift the detrapping peak to lower temperatures, approaching the equilibrium TDS limit in which detrapping and transport are no longer rate-limiting.

    \item In permeation simulations, the effective diffusion coefficient was shown to depend on both the measurement method and the microstructure. The effective diffusion coefficient extracted from the lag time is affected by trap filling, bulk diffusion, and GB diffusion, whereas the effective diffusion coefficient extracted from the steady-state flux is governed primarily by bulk and GB transport.

    \item Oriani effective diffusion coefficient, which accounts for trap filling and bulk diffusion but not connected GB transport, can become inaccurate even for limited GB mobility.

    \item The effective steady-state diffusion coefficient was evaluated over a broad parameter space, and an analytical expression was fitted using non-dimensional variables. This expression provides a practical route to incorporate microstructural parameters, such as grain size, trapping energy, physical GB thickness, and GB mobility, into macroscopic hydrogen transport and embrittlement models.

\end{itemize}

Several limitations of the present model should be noted. The microstructure is assumed to be static, and mechanical effects such as stress-assisted diffusion are neglected. Furthermore, GBs are the only trapping sites considered in the current implementation. The simulations were conducted on single-phase, equiaxed polycrystalline microstructures with uniform GB properties. Nevertheless, the framework is flexible and can be extended to incorporate distinct GB properties, multiphase materials, non-equiaxed microstructures, and additional trapping sites.

\section{Future work}

The proposed framework opens several promising directions for future work. Physically grounded input parameters for GB hydrogen mobility and trapping free energy can be obtained through systematic comparison with atomistic simulations~\cite{Smirnova2023, Hamza2015}. The model can also be extended to account for heterogeneous GB properties~\cite{Hussein2024b, Oudriss2012}, additional trapping sites such as dislocations, vacancies, precipitates, and second-phase interfaces, as well as mechanical effects such as stress-assisted diffusion and hydrogen-stress coupling. In parallel, the framework provides a basis for re-evaluating existing TDS datasets with explicit consideration of microstructural transport pathways~\cite{Oudriss2012}, and for developing combined TDS-permeation methodologies that allow trapping thermodynamics and GB mobility to be identified more independently. These developments can strengthen the link between experimental characterization, microstructure-resolved modeling, and predictive hydrogen embrittlement assessment.

\section*{Data availability}
All codes required to reproduce these findings are available to download from Mendeley data \cite{feyen2026_MendeleyData}.

\section*{Declaration of competing interest}
The authors declare that they have no known competing financial interests or personal relationships that could have appeared to influence the work reported in this paper.

\section*{Declaration of generative AI and AI-assisted technologies in the writing process}
During the preparation of this work, the authors used ChatGPT to improve the readability of the manuscript. After using this tool/service, the authors reviewed and edited the content as needed and take full responsibility for the content of the published article.

\section*{Acknowledgments}
This work was financially supported by the European Union Horizon Europe (HyWay-project, grant agreement number 101135374). The resources and services used in this work were provided by the VSC (Flemish Supercomputer Center), funded by the Research Foundation - Flanders (FWO) and the Flemish Government. The authors thank Ruben Windey, Wout Mertens, and Héléna Verbeeck for their valuable feedback on the manuscript.

% % Appendices require a different numbering of the equations
\renewcommand{\theequation}{A.\arabic{equation}}
\setcounter{equation}{0}
\section*{Appendix A: Numerical evaluation of $C_\mathrm{int}$}
\begin{figure}[pos=h]
	\centering
		\includegraphics[width=0.9\linewidth]{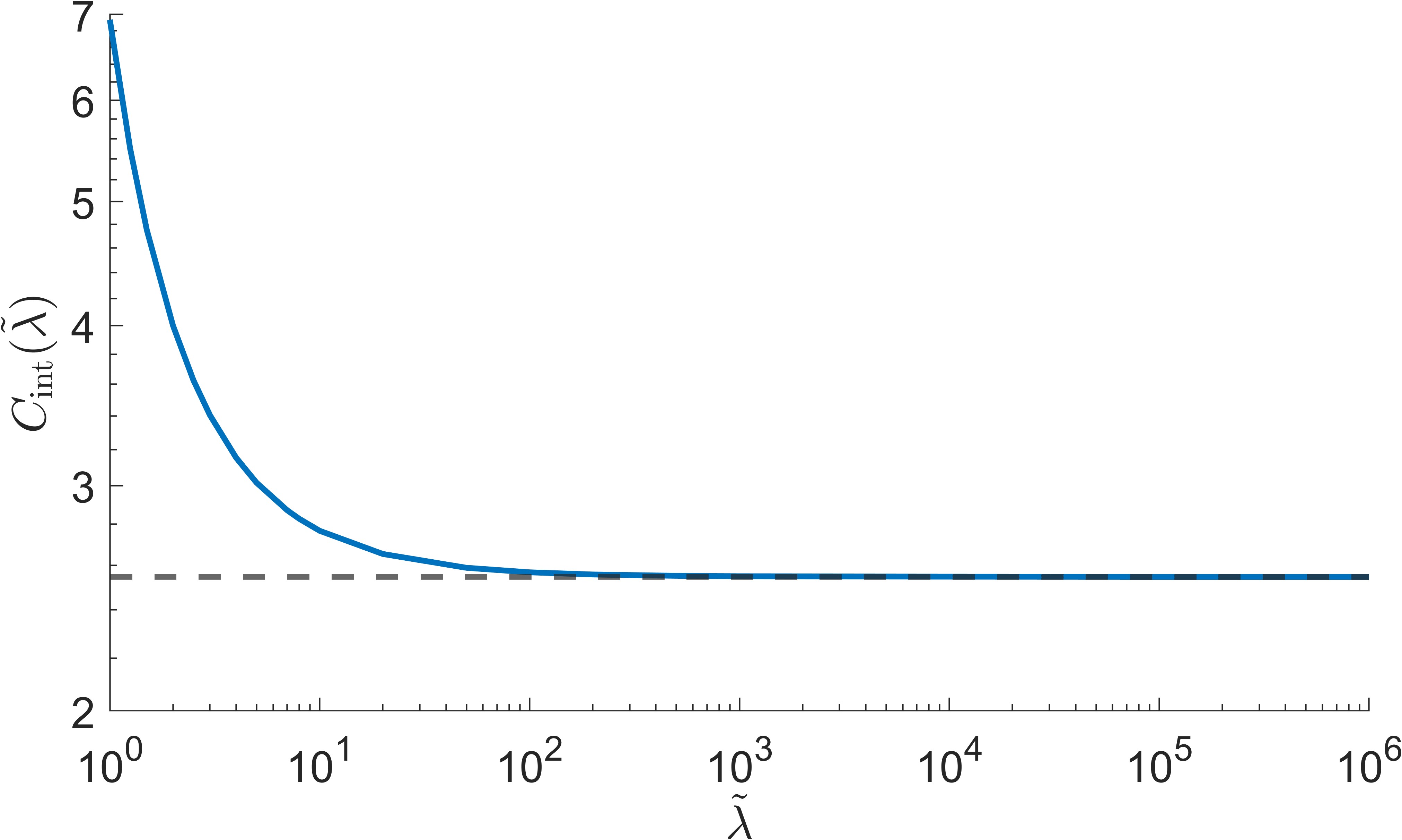}
	\caption{Interfacial constant as a function of the normalized interfacial thickness.}
	\label{FIG:Illustration_C_int}
\end{figure}

To obtain the interface normalization factor $C_\mathrm{int}$, a set of numerical simulations were performed for a system containing a single flat grain boundary, represented by an equilibrium phase-field profile with a hyperbolic tangent shape~\cite{Feyen2023}. The total GB volume was calculated using Equation~\eqref{eq:V_GB}, in which $h_\mathrm{GB}$ was calculated using Equation \eqref{eq:h_GB}, with an initial estimate of \(C_\mathrm{int}\). This value was then iteratively refined until the GB volume obtained by numerical integration matched the physical GB volume, \(A_\mathrm{GB}t_\mathrm{int,phys}\). The calculation was repeated for different normalized interfacial thicknesses ($\tilde{\lambda}$). The resulting values of \(C_\mathrm{int}\) are shown in Figure~\ref{FIG:Illustration_C_int}. The value of \(C_\mathrm{int}(\tilde{\lambda})\) decreases from 6.93 at the true GB scale, \(\tilde{\lambda}=1\), to 2.55 at larger scales. These values are purely geometrical and depend only on the chosen order-parameter profile. In the present work, this profile has a hyperbolic tangent shape, which follows from the double-well potential. Other choices of the phase-field potential, such as a double-obstacle potential, would lead to a different value of \(C_\mathrm{int}(\tilde{\lambda})\). The MATLAB script used for this calculation is available in the supplementary materials~\cite{feyen2026_MendeleyData}.

\renewcommand{\theequation}{B.\arabic{equation}}
\setcounter{equation}{0}
\section*{Appendix B: Numerical solution procedure}
At each time step and grid point of the simulation, the first auxiliary equation for the KKS method, the mass balance Equation (See Equation~\eqref{eq:mass_balance}) and the second auxiliary equation, the local equilibrium condition (see Equation~\eqref{eq:KKS}) need to be solved simultaneously. Using the thermodynamic descriptions given by Equation~\eqref{eq:chempot_B} and \eqref{eq:chempot_GB}, Equation~\eqref{eq:KKS} can be rewritten as:
\begin{equation}
\label{eq:GB_concentration}
\begin{split}
y_\mathrm{H,GB}&= \frac{1 - y_\mathrm{I_2,GB} - ...}{1+\frac{1 - y_\mathrm{H,B} - y_\mathrm{I_2,B} - ...}{y_\mathrm{H,B}} \exp\left(\frac{\Delta \mu_\mathrm{H,b,GB}}{RT}\right)} \\
&= \frac{1 - y_\mathrm{I_2,GB} - ...}{1+\frac{1 - y_\mathrm{H,B} - y_\mathrm{I_2,B} - ...}{y_\mathrm{H,B}}K_\mathrm{GB}}. \\
\end{split}
\end{equation}
Here, $K_\mathrm{GB}$ is the equilibrium constant, determined by the hydrogen binding free energy of the GB and the temperature, introduced here to improve the readability of the equations. It is commonly assumed that both the GB and bulk hydrogen site fractions are small, i.e.,  $y_\mathrm{H,GB} << 1$ and $y_\mathrm{H,B} << 1$. However, this assumption is not generally valid. For example, in the presence of strong trapping, the GB site fraction may approach unity (i.e., $y_\mathrm{H, GB} \to 1 $). For this reason, no low-concentration approximations are made in the present model. 

Substituting Equation~\eqref{eq:GB_concentration} into the mass balance relation (Equation~\eqref{eq:mass_balance}) yields:
\begin{equation}
\label{eq:total_concentration}
\begin{split}
h_\mathrm{B}y_\mathrm{H,B} + \frac{h_\mathrm{GB}\left(1 - y_\mathrm{I_2,GB} - ...\right)}{1+\frac{1 - y_\mathrm{H,B} - y_\mathrm{I_2,B} - ...}{y_\mathrm{H,B}} K_\mathrm{GB}} -y_\mathrm{H} = 0\\
\end{split}
\end{equation}
Solving this nonlinear equation for $y_\mathrm{H,B}$ at each grid point would, in principle, require the use of an iterative solver. In the present work, however, a simpler explicit finite-difference approach based on forward Euler time stepping is employed. Taking the time derivative of Equation   \eqref{eq:mass_balance}, using the definitions of Equation~\eqref{eq:chempot_B} and \eqref{eq:chempot_GB}  and assuming the GBs do not move (i.e. $\partial h_\mathrm{GB}/\partial t = 0$ and $\partial h_\mathrm{B}/\partial t = 0$),  and, for clarity, considering hydrogen as the only interstitial species, leads to:
\begin{equation}
\label{eq:time_derivative_aux}
\resizebox{\linewidth}{!}{$\begin{split}
\frac{\mathrm{d} y_\mathrm{H}}{\mathrm{d} t} =& h_\mathrm{B}\frac{\mathrm{d} y_\mathrm{H,B}}{\mathrm{d} t} + h_\mathrm{GB}\frac{\mathrm{d} y_\mathrm{H,GB}(T,y_\mathrm{H,B})}{\mathrm{d} t}\\
=& h_\mathrm{B}\frac{\mathrm{d} y_\mathrm{H,B}}{\mathrm{d} t} + h_\mathrm{GB}\Bigg(\left(\frac{\partial y_\mathrm{H,GB}}{\partial y_\mathrm{H,B}}\right)_T\frac{\mathrm{d} y_\mathrm{H,B}}{\mathrm{d} t} \\
&+ \left(\frac{\partial y_\mathrm{H,GB}}{\partial K_\mathrm{GB}}\frac{\partial K_\mathrm{GB}}{\partial T}\right)_{y_\mathrm{H,B}}\frac{\mathrm{d}T}{\mathrm{d}t}\Bigg)\\
=& \left(h_\mathrm{B} + h_\mathrm{GB}\frac{K_\mathrm{GB}}{\left(K_\mathrm{GB}(1-y_\mathrm{H,B}) + y_\mathrm{H,B}\right)^2}\right)\frac{\mathrm{d} y_\mathrm{H,B}}{\mathrm{d}t}\\
&- h_\mathrm{GB}\frac{y_\mathrm{H,B}(y_\mathrm{H,B}-1)}{\left(K_\mathrm{GB}(1-y_\mathrm{H,B}) + y_\mathrm{H,B}\right)^2}\frac{\Delta \mu_\mathrm{H,b,GB}}{RT^2} K_\mathrm{GB}\frac{\mathrm{d}T}{\mathrm{d}t}\\
=& F_\mathrm{A}\frac{\mathrm{d} y_\mathrm{H,B}}{\mathrm{d} t}+F_\mathrm{B}\frac{\mathrm{d}T}{\mathrm{d}t}\\
\end{split}$}
\end{equation}
Thus: 
\begin{equation}
\label{eq:dy_B_dt_aux}
\begin{split}
\frac{\mathrm{d} y_\mathrm{H,B}}{\mathrm{d} t} =& \frac{\frac{\mathrm{d} y_\mathrm{H}}{\mathrm{d} t}-F_\mathrm{B}(T,y_\mathrm{H,B},\Delta \mu_\mathrm{H,b,GB})\frac{\mathrm{d}T}{\mathrm{d}t}}{F_\mathrm{A}(T,y_\mathrm{H,B},\Delta \mu_\mathrm{H,b,GB})}
\end{split}
\end{equation}
This expression relates changes in the bulk hydrogen concentration to changes in the total hydrogen concentration. At each time step, the evolution of the total hydrogen site fraction is calculated using the diffusion Equation~\eqref{eq:pf_diffusion}.  Using this value and the local heating rate ($\mathrm{d}T/\mathrm{d}t$), the bulk hydrogen site fraction evolution is computed from Equation~\eqref{eq:dy_B_dt_aux}. Hereafter, the GB hydrogen site fraction is recovered using Equation~\eqref{eq:GB_concentration}.

\renewcommand{\theequation}{C.\arabic{equation}}
\setcounter{equation}{0}
\section*{Appendix C: Grain boundary trapping with perpendicular flux}
\begin{figure}[pos=h]
	\centering
		\includegraphics[width=\linewidth]{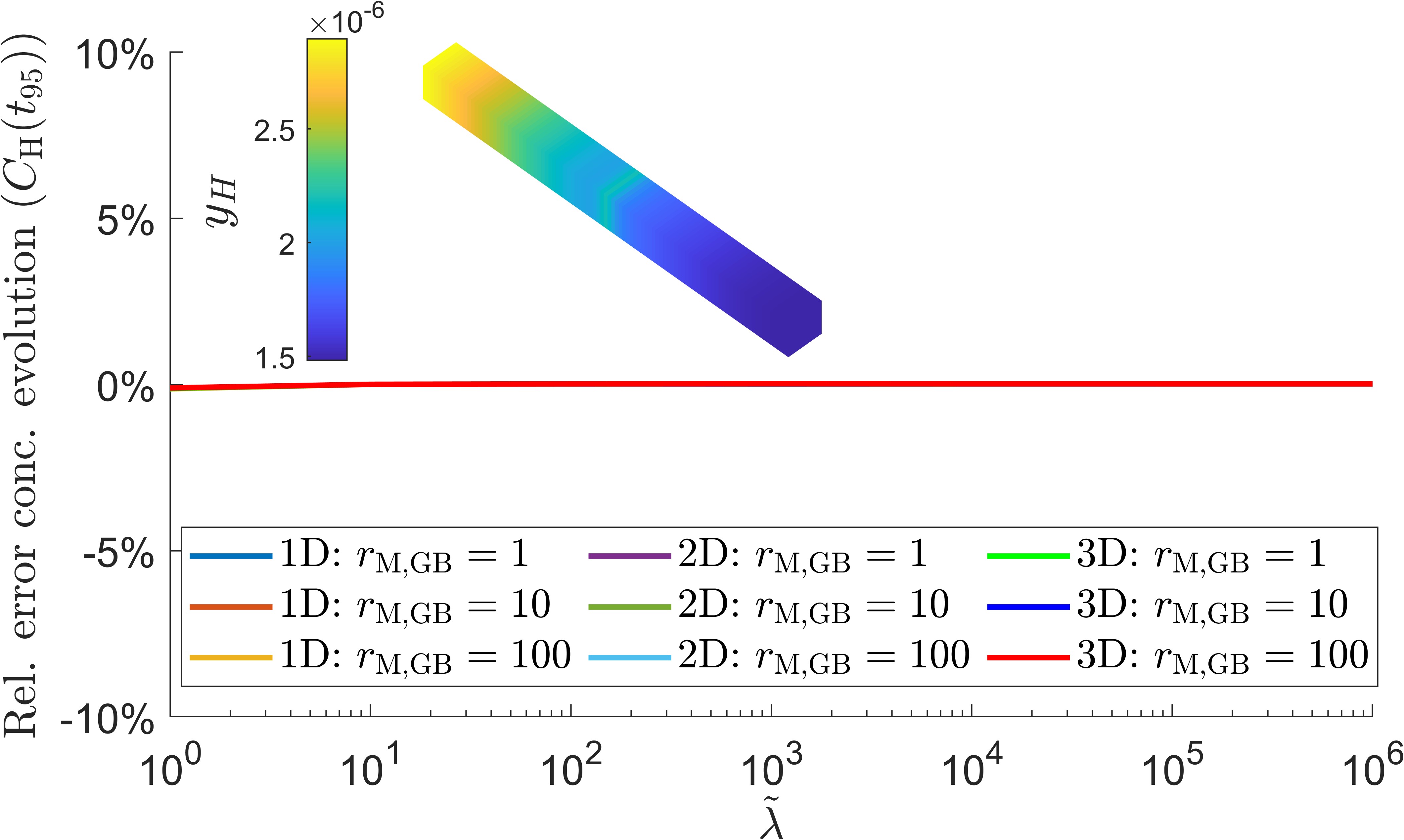}
	\caption{Relative error in the total hydrogen concentration when 95\% of its equilibrium value is reached, compared with the analytical concentration evolution predicted by Fick’s second law, neglecting the finite GB filling time. }
	\label{FIG:BM4}
\end{figure}

 For transport perpendicular to a GB, the kinetics are governed by the imposed boundary conditions and the bulk diffusion coefficient.
In this benchmark, a constant hydrogen concentration $C_0$ is applied at one end of the simulation domain, while Neumann (zero-flux) boundary conditions are imposed at the opposite end. By neglecting the finite time required to fill the GB itself (achieved here by reducing the trapping energy to minimize this contribution, and increasing the size of the simulation in the x-direction), an analytical solution to Fick’s second law can be obtained for comparison \cite{Crank1975}:
\begin{equation}
\label{eq:ficks_second_law_benchmark}
\begin{split}
C =& C_0 - \frac{4C_0}{\pi} \sum_{n=0}^{\infty} \frac{(-1)^n}{2n+1}
\exp\left\{-D(2n+1)^2 \frac{\pi^2 t}{4l^2} \right\}\\
&\cos\left( \frac{(2n+1)\pi x}{2l} \right).\\
\end{split}
\end{equation}
The numerical results are shown in Figure~\ref{FIG:BM4}. 
For all tested spatial dimensions and GB mobilities, the total concentration in the system at the time when 95\% of the final equilibrium concentration is reached, is accurately predicted by the model (all curves overlap). 
Overall, this benchmark confirms that the proposed framework correctly captures the kinetic behavior of hydrogen transport perpendicular to GBs. An analogous analytical solution for the transient filling of a GB by diffusion parallel to the boundary could not be identified; therefore, a corresponding benchmark was not included.

\renewcommand{\theequation}{D.\arabic{equation}}
\setcounter{equation}{0}
\section*{Appendix D: Equilibrium TDS}
This appendix briefly describes the calculation of the equilibrium TDS response. In equilibrium at a given temperature \(T\) and hydrogen partial pressure \(P_\mathrm{H}\), the chemical potential of hydrogen in the gas phase, \(\mu_\mathrm{H,G}(P_\mathrm{H},T)\), is equal to the chemical potential of hydrogen in the bulk phase, \(\mu_\mathrm{H,B}\), given by Equation~\ref{eq:chempot_B}. This condition leads to the following equilibrium bulk lattice site fraction:
\begin{equation}
\begin{split}
y_\mathrm{H,B,e} (P,T)&= \frac{1}{1+\exp\left(\frac{\mu_\mathrm{H,B,0}(T)-\mu_\mathrm{H,G}(P_\mathrm{H},T)}{RT}\right)}. \\
\end{split}
\end{equation}
The expression for \(\mu_\mathrm{H,G}(P_\mathrm{H},T)\) was obtained by a fitting procedure to the TCFE13 database from ThermoCalc \cite{TCFE13}. At equilibrium, the chemical potential of hydrogen in the bulk phase is also equal to the chemical potential of hydrogen in the GB phase, \(\mu_\mathrm{H,GB}\), given by Equation~\eqref{eq:chempot_GB}. This gives the equilibrium GB site fraction:
\begin{equation}
\begin{split}
y_\mathrm{H,GB,e}&= \frac{1}{1+\frac{1-y_\mathrm{H,B,e}}{y_\mathrm{H,B,e}} \exp\left(\frac{\Delta \mu_\mathrm{H,b,GB}}{RT}\right)}.\\
\end{split}
\end{equation}
For an attractive GB trap, \(\Delta \mu_\mathrm{H,b,GB}<0\), so that the GB site fraction is larger than the bulk lattice site fraction. The total amount of hydrogen in the sample (\(N_\mathrm{H}\)), in equilibrium with a given temperature and hydrogen partial pressure, is then given by:
\begin{equation}
\begin{split}
N_\mathrm{H} &=  (V-V_\mathrm{GB})C_\mathrm{H,B} +  V_\mathrm{GB} C_\mathrm{H,GB} \\
&=  \frac{V}{V_\mathrm{m}}f_\mathrm{lattice} \left[\left(1-\frac{V_\mathrm{GB}}{V}\right)y_\mathrm{H,B,e} +  \frac{V_\mathrm{GB}}{V}y_\mathrm{H,GB,e}\right]. \\
\end{split}
\end{equation}
Here $V_\mathrm{GB}$ is the total GB volume. The corresponding equilibrium TDS flux ($J_\mathrm{TDS,eq}$) is obtained from the rate of change of the total hydrogen content during heating. If the desorption flux is defined as positive when hydrogen leaves the sample, this gives:
\begin{equation}
\begin{split}
J_\mathrm{TDS,eq} = -\frac{\partial N_\mathrm{H}(T,P)}{\partial T} \dot{T}. \\
\end{split}
\end{equation}
Here, $\dot{T}$ represents the heating rate during the TDS experiment. In this expression, it is assumed for simplicity that the bulk and GB site fractions are spatially uniform. This approximation corresponds to the equilibrium limit, in which detrapping and transport are sufficiently fast that concentration gradients within the sample remain negligible during heating.

\renewcommand{\theequation}{E.\arabic{equation}}
\setcounter{equation}{0}
\section*{Appendix E: Oriani's equation}
A brief derivation is provided below to illustrate the transformation of Oriani’s original equation into the notation adopted in the present work. The trap-occupancy $ \theta_T$ is equal to the lattice site fraction of hydrogen in the GB, $y_\mathrm{H,GB}$, and the average concentration of trapped and lattice-bound hydrogen ($c_\mathrm{T}$ and $c_\mathrm{L}$) can be obtained via integration of Equation~\eqref{eq:C_H} multiplied by the respective volume fraction ($h_\mathrm{GB}$ and $h_\mathrm{B}$) over the total volume $V$:
\begin{equation}
\begin{split}
D_\mathrm{eff,Or}&= D_L\frac{c_\mathrm{L}}{c_\mathrm{L} + c_\mathrm{T}(1 - \theta_T)} \\
&= D_\mathrm{H,B} \frac{1}{1 + \frac{c_\mathrm{T}}{c_\mathrm{L}}(1 - y_\mathrm{H,GB})}\\
&= D_\mathrm{H,B} \frac{1}{1 + \frac{V_{m}\iiint f_\mathrm{lattice} h_\mathrm{GB} y_\mathrm{H,GB}dV}{V_{m}\iiint f_\mathrm{lattice} h_\mathrm{B} y_\mathrm{H,B}dV}(1 - y_\mathrm{H,GB})}\\
&= D_\mathrm{H,B} \frac{1}{1 + \frac{\iiint h_\mathrm{GB} y_\mathrm{H,GB}dV}{\iiint h_\mathrm{B} y_\mathrm{H,B}dV}(1 - y_\mathrm{H,GB})}\\
&\approx D_\mathrm{H,B} \frac{1}{1 + \frac{y_\mathrm{H,GB}V_\mathrm{GB}}{y_\mathrm{H,B}(V - V_\mathrm{GB})}(1 - y_\mathrm{H,GB})}.\\
\end{split}
\end{equation}
Here $V_\mathrm{GB}$ is the total GB volume. In the final equation, it is assumed for simplicity that the lattice site fractions in the GB and bulk regions are spatially constant. This approximation is justified when only small concentration gradients are present within the system.

\renewcommand{\theequation}{F.\arabic{equation}}
\setcounter{equation}{0}
 \section*{Appendix F: Analytical expression for the effective steady-state hydrogen diffusion coefficient}
\begin{figure}[pos=h]
	\centering
		\includegraphics[width=\linewidth]{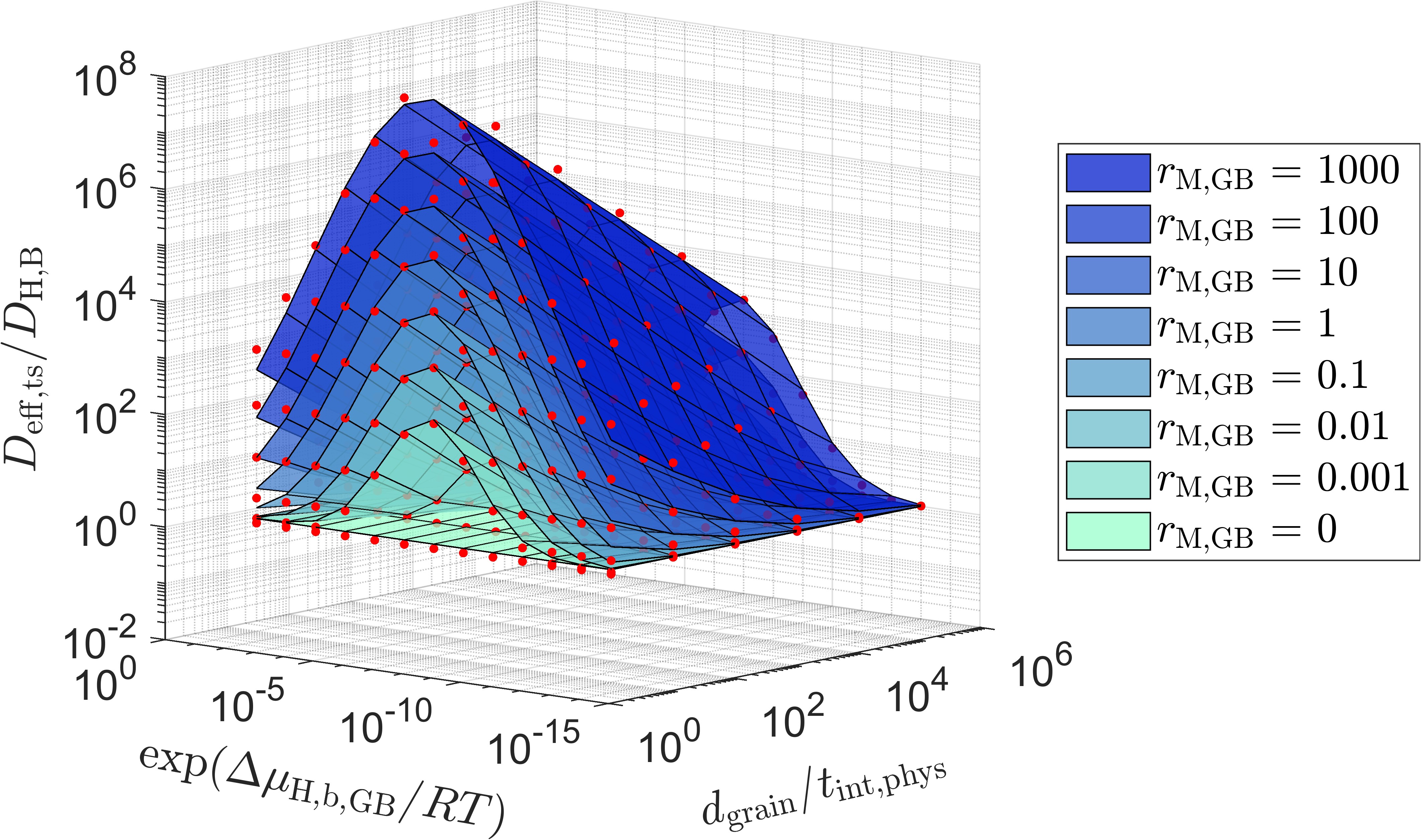}
	\caption{Comparison between the fitted analytical expression (surfaces) and the simulated effective steady-state diffusion coefficients (red dots). }
	\label{FIG:D_eff_fitting}
\end{figure}

 An analytical expression was fitted to the simulated effective hydrogen diffusion coefficient relative to the bulk diffusion coefficient as a function of trapping energy, temperature, average grain size, physical GB thickness, and the ratio between GB and bulk mobility. The resulting formulation is fully dimensionless and is therefore applicable in a general manner across different material systems.
The fitted relationship is given by:
\begin{equation}
\begin{split}
z &= A(x,y) \exp \left(\frac{w - C_4}{C_5 + C_6A(x,y)}\right), \\
\end{split}
\end{equation}
where
\begin{equation}
\begin{split}
A(x,y) &=
\begin{cases}
C_1+C_2x+C_3y, & C_1+C_2x+C_3y>0,\\
0, & C_1+C_2x+C_3y\le 0,
\end{cases}
\end{split}
\end{equation}
with
\begin{equation}
\begin{split}
&w = \log_{10} \left(\exp\left(\frac{\Delta \mu_\mathrm{H,b,GB}}{RT}\right)\right) \\
&x = \log_{10} \left(\frac{d_\mathrm{grain}}{t_\mathrm{int,phys}}\right) \\
&y = \log_{10} \left(\frac{M_\mathrm{at,H,GB,0}}{M_\mathrm{at,H,B,0}}\right) \\
&z = \log_{10} \left(\frac{D_\mathrm{eff,ss}}{D_\mathrm{H,B}}\right). \\
\end{split}
\end{equation}
Here, $d_\mathrm{grain}$ denotes the average grain size, $t_\mathrm{int,phys}$ the physical GB thickness, $M_\mathrm{at,H,GB,0}$ the GB atomic mobility factor, and $M_\mathrm{at,H,B,0}$ the bulk (lattice) mobility factor. This expression only holds for $M_\mathrm{at,H,GB,0}>0$, since $\log_{10}(0)$ is undefined. The fitted coefficients are listed in Table \ref{tab:Coefficients_D_eff_ss} and the results are depicted in Figure~\ref{FIG:D_eff_fitting}. This expression can be directly implemented in macroscopic hydrogen transport and hydrogen embrittlement models, together with Oriani's original equation, to provide microstructure-informed effective diffusion coefficients, offering a quantitative alternative to classical trapping-based formulations. 
\begin{table}[pos=h]
\centering
\caption{Fitted coefficients for the effective steady-state diffusion coefficient. The goodness of fit is quantified by the coefficient of determination, $R^2$, and the root-mean-square error in log-space, \(\mathrm{RMSE}_{\log_{10}}\). Since the fitting was performed in log-space, the \(\mathrm{RMSE}_{\log_{10}}\) value corresponds to a typical multiplicative error of approximately 19\% in the diffusion coefficient, which is acceptable for practical applications.}
\label{tab:Coefficients_D_eff_ss}
\begin{tabular}{@{}lr@{}}
\toprule
Coefficient & Value \\ 
\midrule
$C_1$ & $\,\phantom{-} 5.9998$ \\
$C_2$ & $\,\phantom{-}-0.9351$ \\
$C_3$ & $\,\phantom{-} 0.9471$ \\
$C_4$ & $\,\phantom{-}-6.8222$ \\
$C_5$ & $\,\phantom{-} 0.9005$ \\
$C_6$ & $\,\phantom{-} 0.5898$ \\
\midrule
$R^2$ & $\,\phantom{-} 0.9981$ \\
$\mathrm{RMSE}_{\log_{10}}$ & $\,\phantom{-} 0.0771$ \\
\bottomrule
\end{tabular}
\end{table}

% CRediT authorship
\printcredits

%% Loading bibliography style file
%\bibliographystyle{model1-num-names}
%\bibliographystyle{cas-model2-names}
%\bibliographystyle{ieeetr}
\bibliographystyle{elsarticle-num}

% Loading bibliography database
\bibliography{cas-refs}

%\vskip3pt

\end{document}